\documentclass[twocolumn,preprintnumbers,amsmath,amssymb]{revtex4}

\usepackage{epsfig}
\usepackage{subfigure}
\usepackage{graphics}
\usepackage{amsmath}

\begin{document}

\title{Spectator induced electromagnetic effects in heavy-ion collisions
and space-time-momentum conditions for pion emission}

\author{V.~Ozvenchuk$^1$}%
\email{Vitalii.Ozvenchuk@ifj.edu.pl}%
\author{~A.~Rybicki$^1$}%
\author{~A.~Szczurek$^{1,2}$}%
\author{~A.~Marcinek$^1$}%
\author{~M.~Kie\l{}bowicz$^1$}%
\affiliation{%
$^1$ H.~Niewodnicza\'nski Institute of Nuclear Physics, Polish Academy of Sciences, %
 31-342 Krak\'ow, Poland\\
$^2$ University of Rzesz\'ow, %
 35-959 Rzesz\'ow, Poland %
}

\date{\today}

\begin{abstract}
We present our calculation of electromagnetic effects, induced by
the spectator charge on Feynman-$x_F$ distributions of charged pions
in peripheral $Pb+Pb$ collisions at CERN SPS energies, including
realistic initial space-time-momentum conditions for pion emission.
The calculation is performed in the framework of
our simplified implementation of the fire-streak
model, adopted to the production of both $\pi^-$ and $\pi^+$ mesons.
Isospin effects are included to take into account the asymmetry in
production of $\pi^+$ and $\pi^-$ at high rapidity. A comparison to
a {simpler} model from the literature is made. We obtain a good
description of the NA49 data on the $x_F$- and $p_T$-dependence of
the ratio of cross sections $\pi^+/\pi^-$. The experimental data
favors short times ($0.5<\tau<2$~fm/$c$) for fast pion creation in
the local fire-streak rest frame. The possibility of the expansion
of the spectators is considered in our calculation, and its
influence on the electromagnetic effect observed for the
$\pi^+/\pi^-$ ratio is discussed.
In addition we discuss the relation between anisotropic flow and
the electromagnetic distortion of $\pi^+/\pi^-$ ratios,
and study the influence of transverse expansion of fire streaks as well
as their vorticity on this distortion.
In this latter study we find that inclusion of
%includes the connection between
rotation of fire streaks in our model gives a satisfactory description
of the rapidity dependence of pion directed flow.
We conclude that {our implementation of the} fire-streak
model, which properly describes the centrality dependence of $\pi^-$
rapidity spectra at CERN SPS energies, also provides
%realistic initial conditions for pion production. Consequently, it provides
a quantitative description of the electromagnetic effect on the
$\pi^+/\pi^-$ ratio as a function of $x_F$.
\end{abstract}

\maketitle

%-----------------------
\section{Introduction}
\label{introduction}
%-----------------------
More than ten years ago~\cite{EM_previous_1} two of us presented
model calculations on a somewhat spectacular electromagnetic effect
caused by fast moving spectators on charged pion momentum
distributions at CERN Super Proton Synchrotron (SPS) energies. The
effect is most peculiar in Feynman-$x_F$ distributions of $\pi^+$
and $\pi^-$ when limiting to low pion transverse momenta. This
phenomenon, observed in the NA49
experiment~\cite{NA49_EM,NA49_EM_2}, was explained in a simple toy
model where a point-like source of pions was assumed. The only free
parameter of the model was the distance between the source and the
spectator system, which in the present paper we will refer to as
$d_E$. After the initial emission of $\pi^+$ and $\pi^-$, the
calculation  of their trajectories in the electromagnetic field of
the two spectator systems was performed. A good description of the
data was obtained when the original pion source was not far from the
spectator. The best agreement with the experimental
data~\cite{NA49_EM} was obtained for $d_E\approx0.5-1$~fm. The
interpretation of this fact was not given in~\cite{EM_previous_1}.
The simplest explanation could be a fast hadronization of the
plasma, which seems difficult to reconcile with the present
knowledge on pion decoupling times at mid-rapidity~\cite{hbt}.
Definitely a deeper conclusion was not possible within the toy model
considered in~\cite{EM_previous_1}. The same simple model was able
to describe~\cite{EM_previous_2,RS2014} a rather small effect of
splitting of directed flow ($v_1$) for $\pi^+$ and $\pi^-$ as
observed at RHIC~\cite{STAR_v1}.

The old version of the model was rather simplistic. The question
arises whether models with more realistic initial conditions can
describe the electromagnetic effects observed for $\pi^+$ and
$\pi^-$ spectra. Recently it was shown that one can describe the
broadening of the $\pi^-$ rapidity distribution with centrality, as
observed by the NA49 experiment~\cite{NA49_rapidity}, within a
special {(simplified)} implementation of the fire-streak model
{which we}
proposed in
Ref.~\cite{FS_model_1}. In our opinion the fire-streak model (see
also Refs.~\cite{fs1,fs2,fs3,fs4,fs5,fs6,fs7}) provides realistic
initial conditions for quark-gluon plasma creation at SPS energies,
in particular obeying energy-momentum conservation. In this model,
for peripheral collisions, the initial quark-gluon plasma moves with
different velocities as a function of the impact parameter vector
$(b_x, b_y)$ (see Fig.~\ref{schematic_collision}). Do
electromagnetic effects, seen in the $\pi^+ / \pi^-$ ratios, support
such a picture? We shall try to answer this question in the present
paper. We consider also simplified initial conditions to investigate
to what extent the fire-streak model with local energy-momentum
conservation provides more realistic initial conditions to
understand experimental results for the $\pi^+ / \pi^-$ ratio.

With the extended pion source, it seems difficult a priori to
describe the data if different emission zones do not cooperate in a
proper (generally unknown) way. Having the extended source in the
present calculation, we wish to understand in addition the success
of the simplified calculation with the point-like source
\cite{EM_previous_1}. We wish to address also the issue how
important is the {\it tilted condition} ($v_z = v_z(b_x,b_y)$) for
describing the electromagnetic effect.

This paper is organized as follows.
In Sec.~\ref{general} we present a few general remarks on our approach.
In Sec.~\ref{initial_conditions}
we provide a detailed description of the initial conditions for pion
creation. We then discuss in Sec.~\ref{event_generator} the event
generator implemented in our model. In Sec.~\ref{results} we present
the results of the calculation of electromagnetic effects on the
$x_F$-distribution of charged pions within the fire-streak model.
The summary and conclusions are given in Sec.~\ref{summary}.

%-------------------------------------------------------------------------
\section{General remarks about theoretical approach}
\label{general}
%-------------------------------------------------------------------------

The details of the electromagnetic effect were presented in
Ref.~\cite{EM_previous_1} whereas the details about our
version of the fire-streak model in Refs.~\cite{FS_model_1,FS_model_2}.
We will not repeat these details here.
%
%%%
However, we point out that in spite of the basic similarity of our
approach formulated in Refs.~\cite{FS_model_1,FS_model_2}
to the original fire-streak concept of
Refs.~\cite{fs1,fs2,fs3,fs4,fs5,fs6,fs7},
differences exist on the more detailed level. These become clearly
apparent from the comparison of the cited works.
%%%

Our fire-streak model, in contrast to other models in the literature,
has a rather small number of parameters, essentially only three free
parameters for the so-called fire-streak fragmentation function \cite{FS_model_2}.
It describes nicely the rapidity distributions of pions
as well as the broadening in rapidity distributions when going from central
to peripheral collisions which is due to local energy-momentum conservation
within a given fire streak~\cite{FS_model_1}.
%In contrast to other models in the literature is gives realistic space-time initial conditions for plasma evolution.

As will be discussed below, the evolution of charged pions in the
electromagnetic (EM) field of spectators {requires} long times. The
long times of the EM evolution require to use a simple treatment of
the plasma stage. Our fire-streak model has all necessary features.
It is relatively simple, and describes the pion production at the
SPS energies. In future one could think to use a 3D hydrodynamical
program but {this} clearly goes beyond the present interest.

Instead, in the present analysis we will include {anisotropic} flow effects,
thought to be of hydrodynamical origin, in a more phenomenological way
-- as {an} initial condition for the EM evolution.
The initial conditions will be discussed in detail in Sec.~\ref{initial_conditions} below.

Finally, we note that plasma in the $Pb+Pb$ collision
at SPS energies is not isospin ($u-d$ or $\bar u - \bar d$)
symmetric. The asymmetry in quark/antiquark production is often called
isospin effect.
We will include the isospin effects due to initial quark asymmetry
in a simplified, yet realistic way.

%Below we shall discuss in detail initial conditions.

%----------------------------------------------------------------
\section{Initial conditions for pion creation}
\label{initial_conditions}
%----------------------------------------------------------------
In this section we discuss the initial conditions for pion creation
which were implemented into our approach.

%----------------------------------------------------------------------
\begin{figure}
\includegraphics[width=0.5\textwidth]{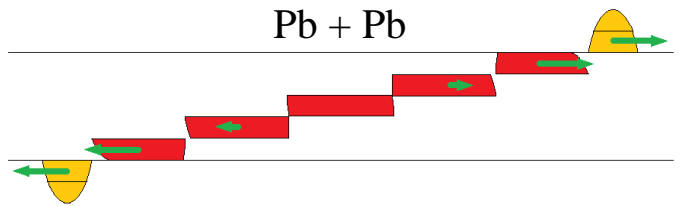}
\vspace{0.0cm} \caption{The situation after the collision. The area
marked in red shows the partonic matter. Each element moves with a
different longitudinal velocity, which can be obtained from
energy-momentum
conservation~\cite{FS_model_1}.}\label{schematic_collision}
\end{figure}
%-----------------------------------------------------------------------

%----------------------------------------------------------
\subsection{Rapidity distribution of pions}
\begin{figure}
\includegraphics[width=0.5\textwidth]{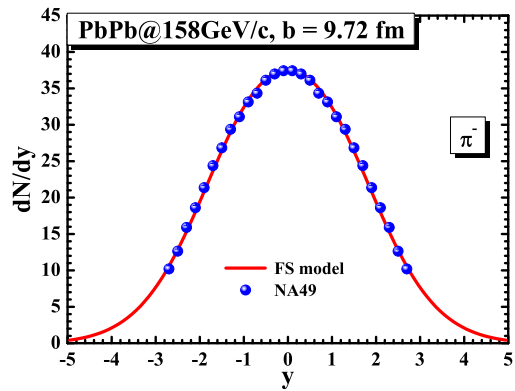}
\vspace{-1.9cm} \caption{The rapidity distribution of negative pions
obtained from the fire-streak model (solid red line) in comparison
to the NA49 experimental data~\cite{NA49_rapidity} (blue symbols)
for peripheral $Pb+Pb$ collisions at top SPS
energy.}\label{dNdy_comparison}
\end{figure}

%----------------------------------------------------------
\label{Rap-distr-pions} As was mentioned in Sec.~\ref{introduction}
we previously studied the electromagnetic effects of the spectator
charge on the momentum spectra of $\pi^+$ and $\pi^-$ produced in
peripheral $Pb+Pb$ collisions at SPS
energies~\cite{EM_previous_1,EM_previous_2} using a point-like
source for pion creation. In the present paper we use the initial
(without electromagnetic effects included) rapidity distribution of
negative pions~\footnote{The rapidity distribution of $\pi^+$ was
not measured for the $Pb+Pb$ collisions at SPS
energies.} obtained from our version of
the fire-streak model formulated in
Refs.~\cite{FS_model_1,FS_model_2}. This model well describes the
rapidity distribution of $\pi^-$ in comparison to the NA49
experimental data~\cite{NA49_rapidity}, which is shown in
Fig.~\ref{dNdy_comparison} for the most peripheral collisions.
%
%----------------------------------------------------------------------
%-----------------------------------------------------------------------
%
The fire-streak fragmentation function into negative pions was
parametrized in Ref.~\cite{FS_model_1} in the form:
\begin{equation}
%\frac{dn}{dy}(y,y_s,E_s^*,m_s)=A\cdot(E_s^*-m_s)\cdot\exp\biggl(-\frac{[(y-y_s)^2+\epsilon^2]^{\frac{r}{2}}}{r\sigma_y^r}\biggr).
\frac{dn}{dy}(y,y_s,E_s^*,m_s)\!=\!A(E_s^*-m_s)\exp\!\biggl(\!-\frac{[(y-y_s)^2+\epsilon^2]^{\frac{r}{2}}}{r\sigma_y^r}\!\biggr),
\label{fragmentation_function}
\end{equation}
where $y$ is the rapidity of the pion, $y_s$ is the fire-streak
rapidity given by energy-momentum conservation, $E_s^*$ is its total
energy in its own rest frame, and $m_s$ is the sum of ``cold'' rest
masses of the two nuclear ``bricks'' forming the fire streak (see
Ref.~\cite{FS_model_1}). The free parameters of the
function~(\ref{fragmentation_function}) are $A$, $\sigma_y$ and $r$,
which appeared common to all fire streaks and independent of $Pb+Pb$
collision centrality. The fit of the NA49 centrality selected
$Pb+Pb$ data~\cite{NA49_rapidity} gave $A=0.05598$,
$\sigma_y=1.475$, and $r=2.55$. Finally, $\epsilon$ is a small
number ensuring the continuity of derivatives ($\epsilon=0.01$ was
used in Ref.~\cite{FS_model_1}). The
expression~(\ref{fragmentation_function}) defines the distribution
of negative pions created by the fragmentation of a single
fire streak. The resulting pion spectrum in a given centrality was
constructed as the sum of independent rapidity fragmentation
functions:
\begin{equation}
\frac{dn}{dy}(y,b)=\sum_{(i,j)}\frac{dn}{dy}\bigl(y,y_{s_{(i,j)}}(b),E^*_{s_{(i,j)}}(b),m_{s_{(i,j)}}(b)\bigl),
\label{pion_rapidity}
\end{equation}
where $(i,j)$ denominate the position of a given fire streak in the
transverse $(x,y)$ plane, and $b$ is the impact parameter of the
$Pb+Pb$ collision.

%---------------------------------------------------------------------------
\subsection{Transverse-momentum distribution of pions}
%---------------------------------------------------------------------------
\begin{figure}
%\vspace{0.1cm}
\includegraphics[width=0.5\textwidth]{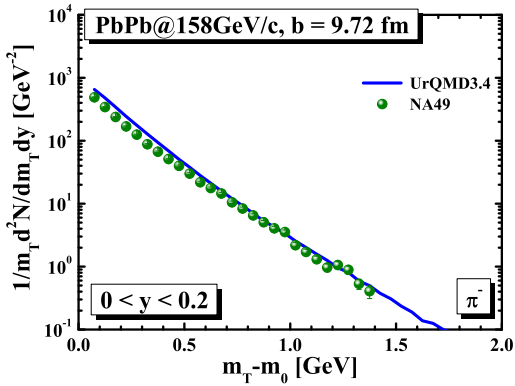}
\vspace{-1.9cm} \caption{The UrQMD v3.4 predictions (solid blue
line) for the transverse mass spectrum at midrapidity $(0<y<0.2)$ of
$\pi^-$ produced in peripheral $Pb+Pb$ collisions at 158 GeV/$c$ in
comparison to the experimental data from the NA49 Collaboration
(green symbols)~\cite{NA49_rapidity}.} \label{dNmtdmtdy_comparison}
\end{figure}
%---------------------------------------------------------------------------
In Ref.~\cite{FS_model_1} only rapidity distributions of pions were
studied. For the discussion of electromagnetic effects we are
interested also in transverse-momentum distributions of pions.

%----------------------------------------------------------------------------
\begin{figure*}
\vspace{0.3cm} \centering \subfigure{
\resizebox{0.47\textwidth}{!}{%
 \includegraphics{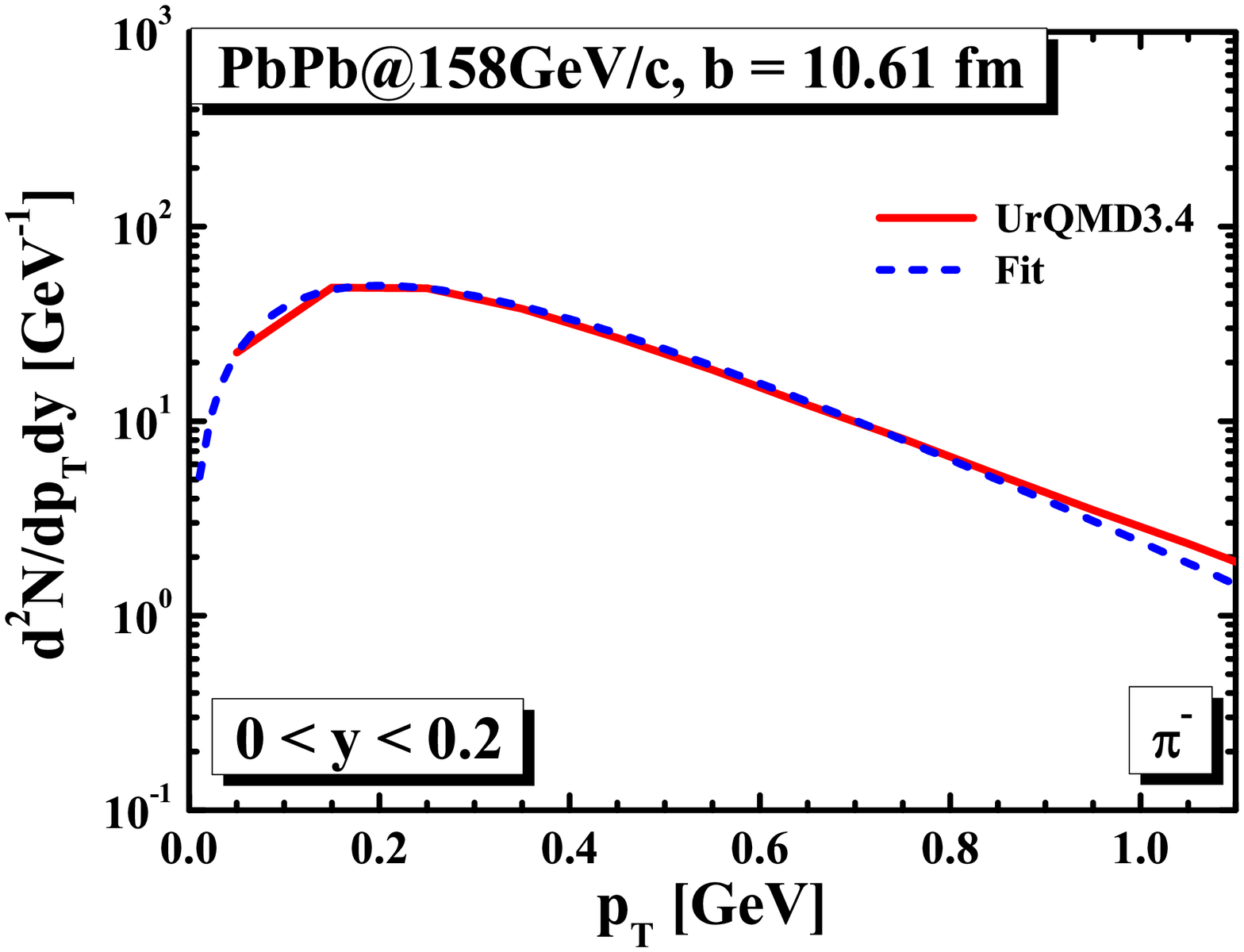}\hspace*{-5cm}
} } \subfigure{
\resizebox{0.47\textwidth}{!}{%
 \includegraphics{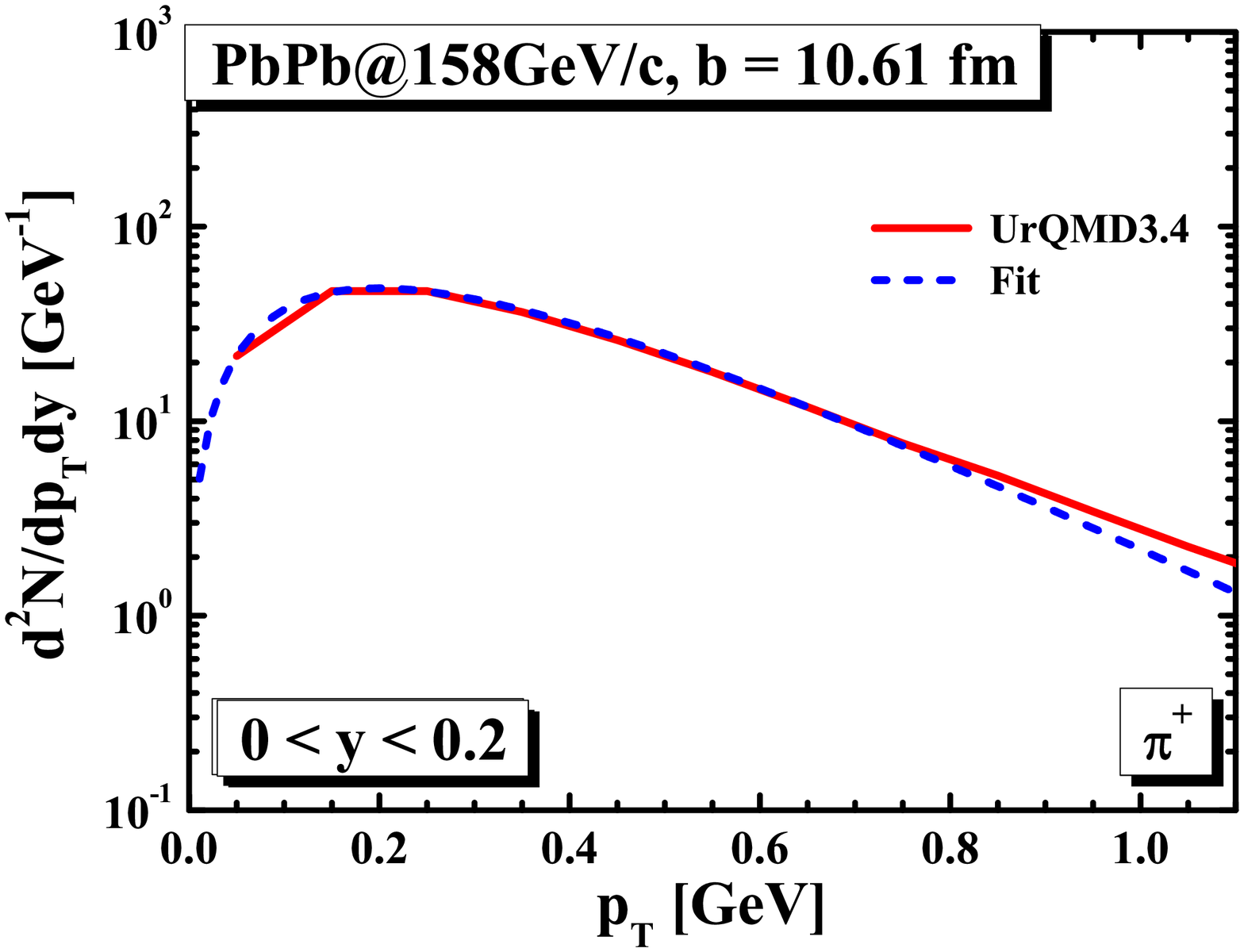}\hspace*{-5cm}
} } \vspace{-1.7cm}\caption{The UrQMD v3.4 simulations (solid red
lines) of transverse-momentum spectra at midrapidity of $\pi^-$
mesons (left) and of $\pi^+$ mesons (right) produced in peripheral
$Pb+Pb$ collisions at 158 GeV/$c$ beam momentum. The dashed blue
lines represent the corresponding fits, made according to
Eq.~(\ref{fit_function}).} \label{fit}
\end{figure*}
%----------------------------------------------------------------------------

For the initial transverse-momentum distribution of pions we choose
the one obtained from the UrQMD 3.4
simulations~\cite{UrQMD_1,UrQMD_2}. In
Fig.~\ref{dNmtdmtdy_comparison} we compare the UrQMD 3.4 predictions
with the NA49 data for the transverse-mass spectrum at midrapidity
of negative pions produced in peripheral $Pb+Pb$ collisions at top
SPS energy. The model reasonably well describes the experimental
data.

We parametrize the resulting UrQMD 3.4 predictions for
transverse-momentum distributions of pions by the exponential
function~\cite{fit_1,fit_2}:
\begin{equation}
\label{fit_function}
\frac{dN}{dp_T}=\frac{Sp_T}{T^2+mT}\exp{[-(m_T-m)/T]},
\end{equation}
where $m$ is the mass of the pion, $m_T=\sqrt{m^2+p_T^2}$ is its
transverse mass, $S$ and $T$ are the yield integral and the inverse
slope parameter, respectively. The transverse-momentum distributions
of pions are normalized as follows
\begin{equation}
\int_{0}^{\infty}\frac{dN}{dp_T}dp_T=S. \label{normalization}
\end{equation}

The fit to the UrQMD v3.4 simulations at midrapidity is presented in
Fig.~\ref{fit} and it gives $T_{\pi^-}=165$~MeV and
$T_{\pi^+}=163$~MeV. In general, the inverse slope parameter may
depend on rapidity, $T = T(y)$. However, we performed the study on
rapidity dependence of the inverse slope parameter by fitting the
UrQMD results for different rapidity bins and extracting the inverse
slope parameters for the corresponding rapidity. We figured out that
the results of calculation of electromagnetic effects on the
$x_F$-distribution of charged pions do not depend much when changing
the inverse slope parameter with the pion rapidity. Therefore, for
further discussion we assume that the inverse slope parameter does
not depend on rapidity and equals to its midrapidity value mentioned
above.

%--------------------------------
{\subsection{Particle flow}
%--------------------------------

In the present paper we wish to quantify (for the first time) the
possible effect of pion
%the particle
flow on the observed $\pi^+ / \pi^-$
ratio. The particle flow is a well established empirical and
phenomenological fact observed at SPS, RHIC and LHC energies.
The flow is quantified in terms of the Fourier decomposition:
\begin{eqnarray}
&&\frac{d N(y,p_T,\phi)}{d p_T d \phi} = \frac{d N(y,p_T)}{d p_T} \nonumber \\
&&\left( 1 + 2 v_1(y,p_T) cos( \phi) + 2 v_2(y,p_T) cos(2 \phi) +
...\right) \label{flow_decomposition}
\end{eqnarray}
for different particle species. The $\phi$ angle above is the
azimuthal angle with respect to the reaction plane, in our case the
plane spanned by the noncentral trajectories of the colliding nuclei
(or outgoing spectators). The flow coefficients depend on collision
energy, rapidity and transverse momentum. The symmetry dictated by
the geometry imposes $v_2, v_1 \to$ 0 when $p_T \to$ 0. The
electromagnetic
effect
%distortion of the $\pi^+ / \pi^-$ ratio
is concentrated at rather small transverse
momenta~\cite{EM_previous_1} so naively one could expect no significant
influence from flow.
In addition $v_2(-y) = v_2(y)$ and $v_1(-y) = -v_1(y)$.

The NA49 collaboration clearly observed \cite{NA49_flow} nonzero
elliptic ($v_2$) and directed ($v_1$) pion flow coefficients. The nonzero
flow coefficients are usually interpreted in terms of hydrodynamical
evolution of the QGP fluid. A good description of the experimental
data is obtained at midrapidities. Here we are interested rather in
very forward rapidities (large $x_F$) where there is no general
consensus on the underlying dynamics. Therefore in the following
analysis we prefer rather to parametrize experimental data on $v_2$
and $v_1$ as a function of pion rapidity and transverse momentum.
In Fig.~\ref{fig:flow_fit} we show our fit to the NA49 data
\cite{NA49_flow}. We present results for both elliptic $v_2$ (upper
row) and directed $v_1$ (lower row). Our purely mathematical fit
nicely represents the NA49 experimental data \cite{NA49_flow} in a
broad range of rapidities and transverse momenta. For our
electromagnetic effect we need data in a broad range of $x_F$, i.e.
at rather large rapidities. Therefore we included also the data of
the WA98 collaboration \cite{WA98_flow} {\rm into the fit}.
%Our mathematical fit seems to be a good representation of the NA49 data.

%----------------------------------------------------------------------
\begin{figure}
\includegraphics[width=4.2cm]{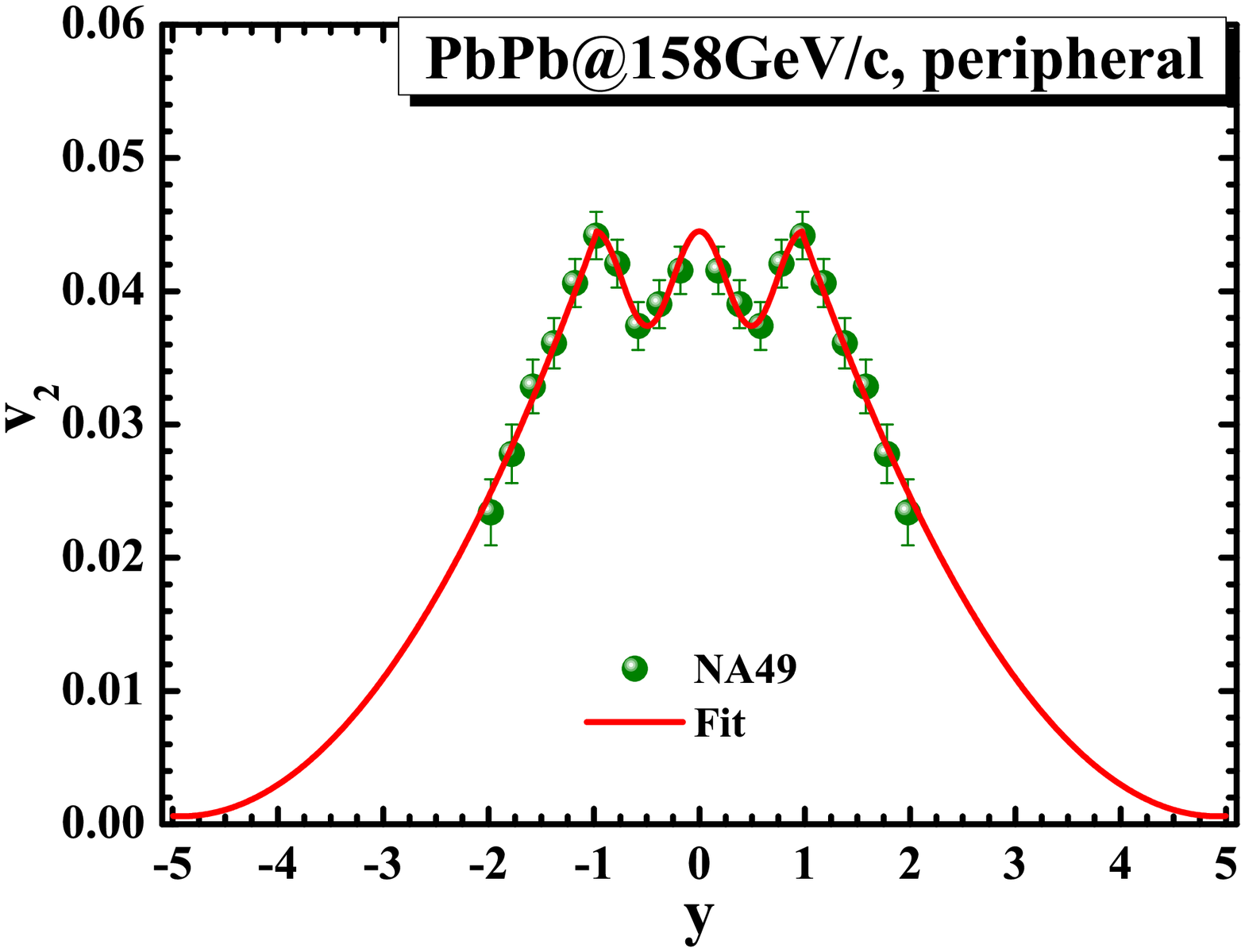}
\includegraphics[width=4.2cm]{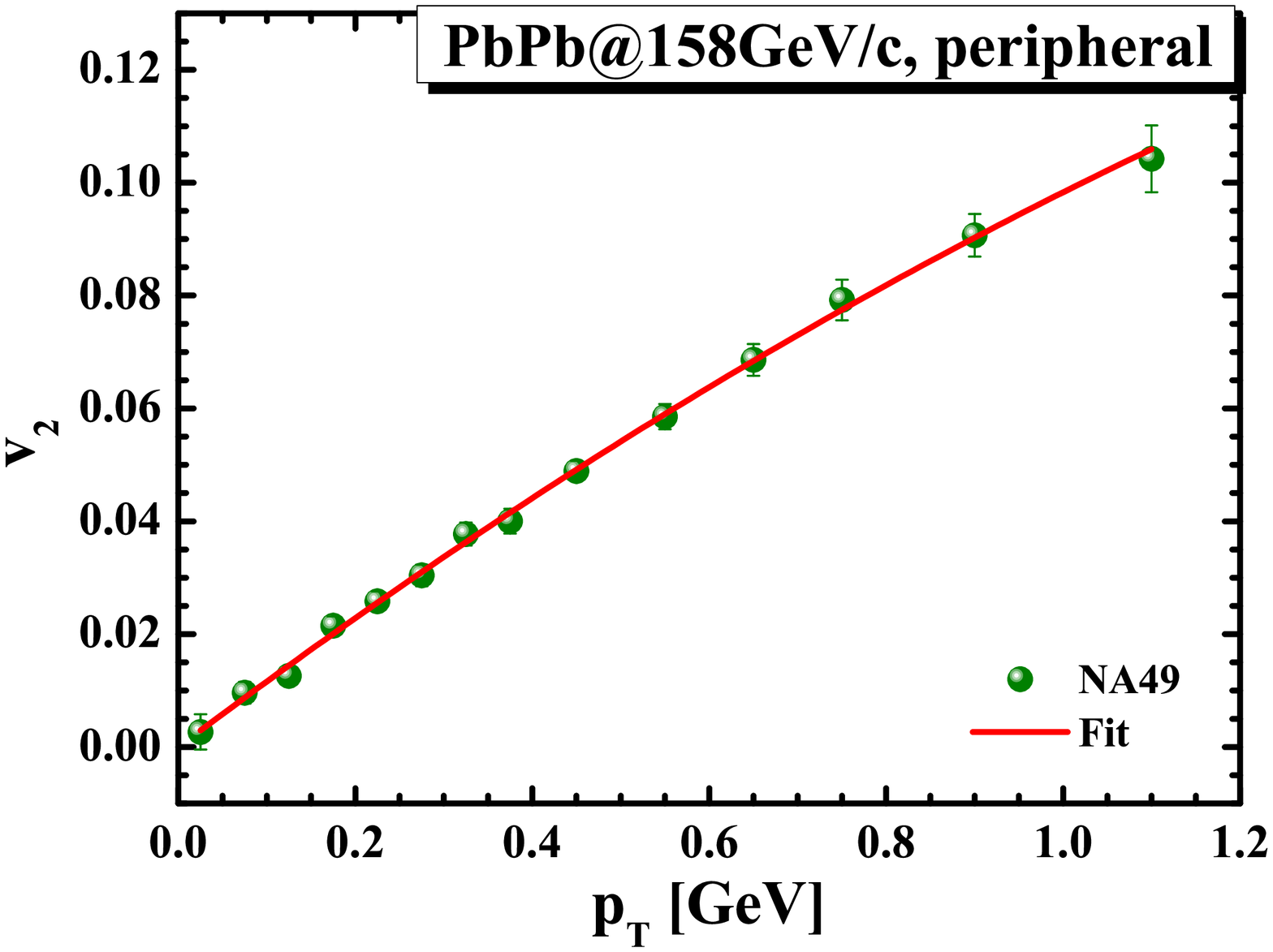}
\includegraphics[width=4.2cm]{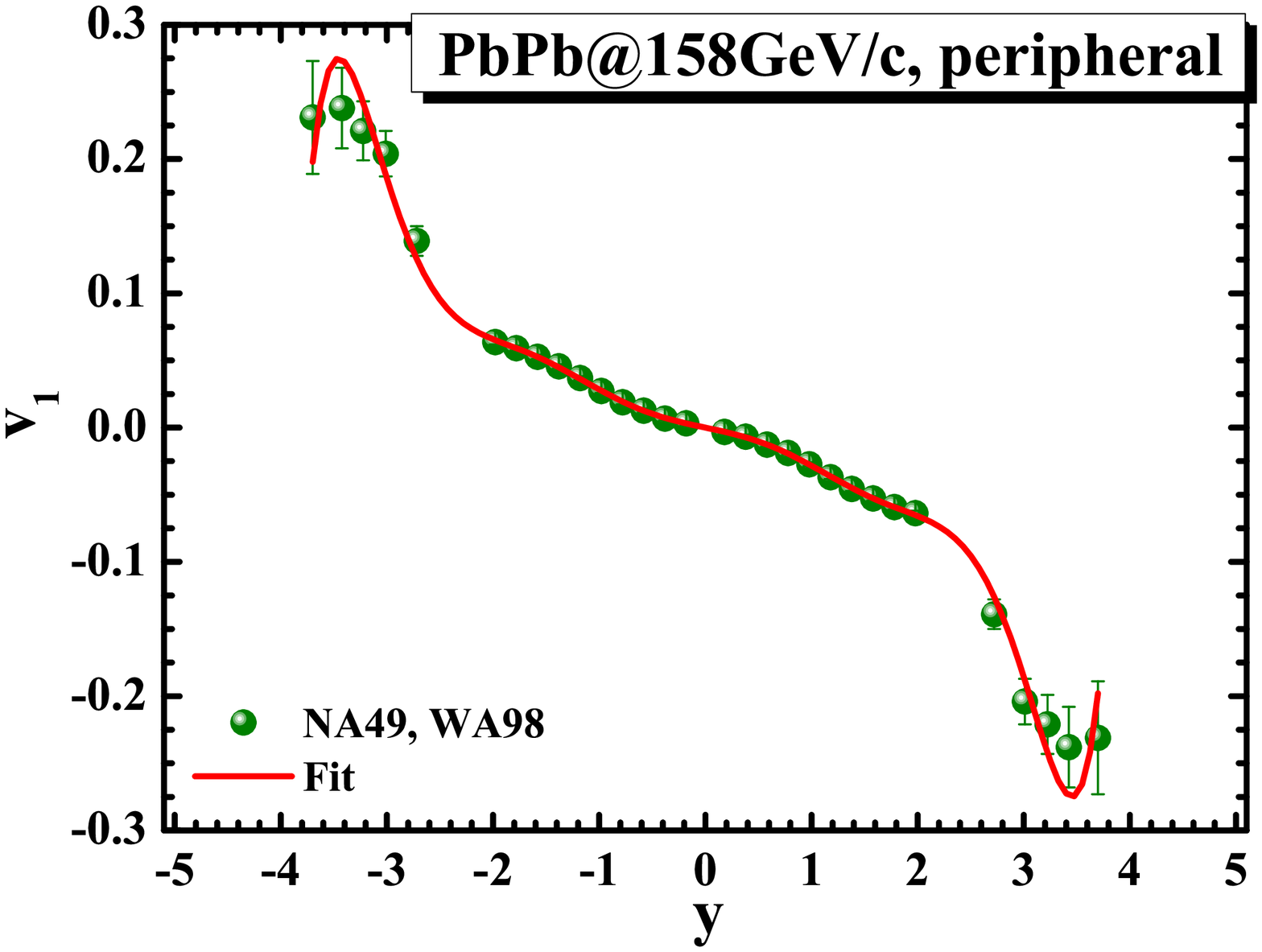}
\includegraphics[width=4.2cm]{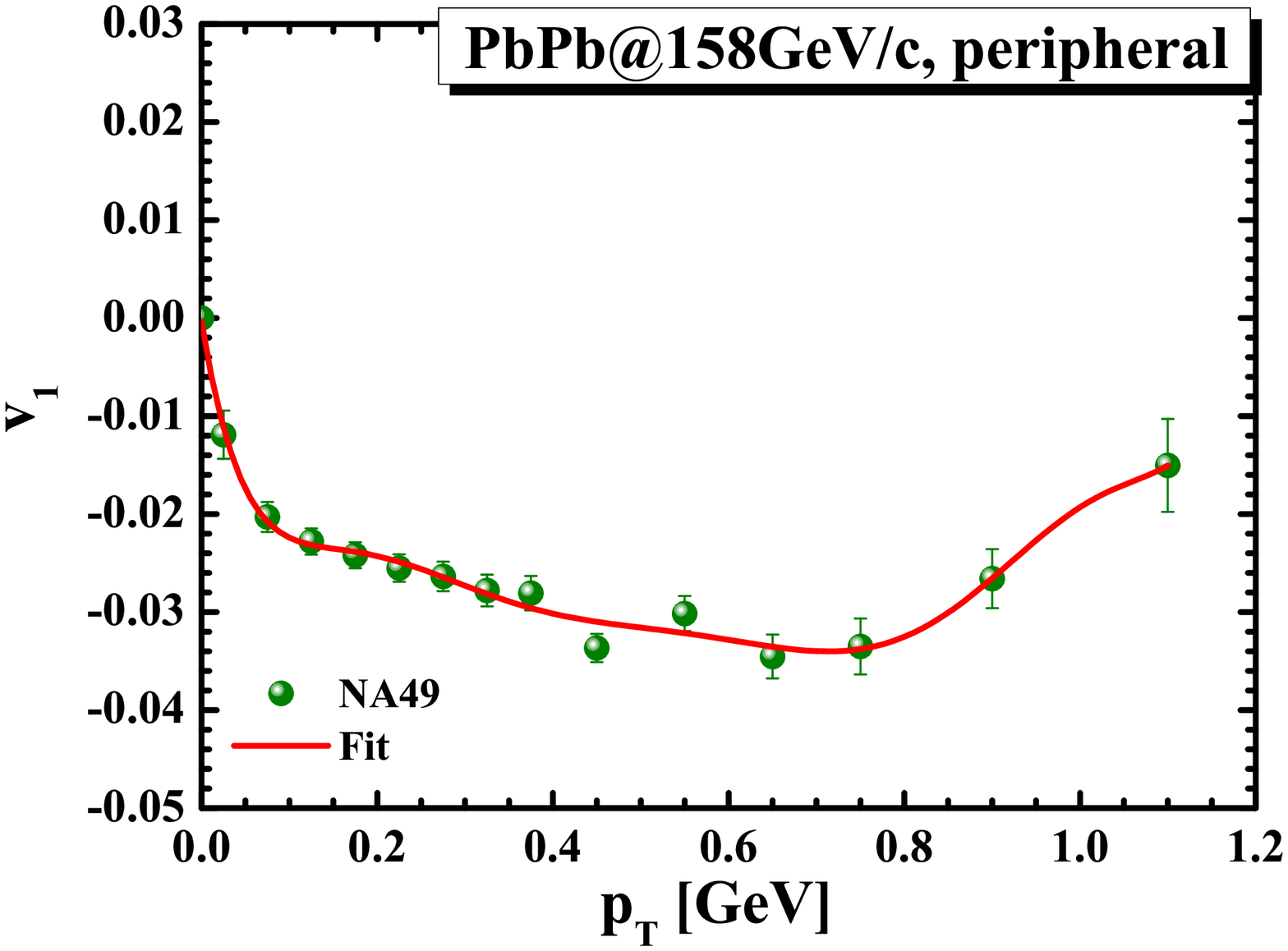}
\caption{ Our fit to average $v_2$ (upper row) and $v_1$ (lower row)
data obtained by the NA49 {and WA98 collaborations}. The left panels
are for rapidity dependence while the right panels are for
transverse momentum dependence. } \label{fig:flow_fit}
\end{figure}
%-----------------------------------------------------------------------

In our calculation on electromagnetic evolution of pions we need to
have $v_{1,2} = v_{1,2}(y,p_T)$, i.e. dependence on two variables
$y$ and $p_T$ simultaneously. In the following we assume:
\begin{equation}
v_{1,2}(y,p_T) = v_{1,2}(y) \times v_{1,2}(p_T) / v_{1,2}^{eff},
\label{two-dim_parametrizations}
\end{equation}
where $v_{1,2}^{eff}$  are effective parameters adjusted in order to
nicely reproduce the separate dependences of $v_1$ and $v_2$ on $y$
and $p_T$ within such a factorized ansatz in our EM code.

%----------------------------------------
\subsection{Pion emission time}
%----------------------------------------
To start the calculation of electromagnetic effects we have to fix
also the time of emission of pions from the fire streaks (which we
consider as a longitudinally expanding plasma), and the
corresponding initial position of the pion relative to the two
spectator systems. This cannot be calculated from first principles
and it is treated here as a free parameter. We assume the time of
creation of pions, $\tau$, in the fire-streak rest frame. Up to this
time the fire streak evolves in the longitudinal direction. For our
calculations of electromagnetic effect we have to calculate the
actual position of pion creation in $z$ for each ($i,j$) fire streak
in the nucleus-nucleus center-of-mass system. This can be done by
applying the Lorentz transformation
\begin{equation}
\tau \to t_{i,j} \; . \label{Lorentz_transformation}
\end{equation}
The transformation is given by the velocity of a given fire streak
in the center-of-mass system which in turn is given, in our model,
by its position in the impact parameter plane
$(b_x,b_y)$~\cite{FS_model_1}.

%-----------------------
\section{Event generator for the whole reaction}
\label{event_generator}
%-----------------------
Having fixed the position of the pion and its momentum vector we can
start the evolution of the charged pion trajectory in the
electromagnetic field of both spectators using a Lorentz invariant
formalism \cite{EM_previous_1}. This evolution requires rather long
evolution times. In our calculation we use $t_{max}$ of the order of
$10^4$~fm/$c$. We note that this time is taken in the
nucleus-nucleus center-of-mass system.
%
%{\it Our evolution of charged pions in the electromagnetic field ofspectators requires very long times.
In Fig.~\ref{fig:ratio_tmax} we show the $\pi^+/\pi^-$ ratio
for different evolution times $t_{max}$ = $10^2$, $10^3$, and
$10^4$ fm/c. We see a convergence of the electromagnetically
deformed $\pi^+ / \pi^-$ ratio only for $t_{max} > 10^3$ fm/c.
Such long times are required because the emitted pions of interest
in the EM ``dip'' have velocities similar to spectators, i.e. stay
close to spectators for fairly long time. For $\pi^-$ mesons even an
orbitting is possible \cite{EM_previous_1}.

%----------------------------------------------------------------------
\begin{figure}
\includegraphics[width=0.5\textwidth]{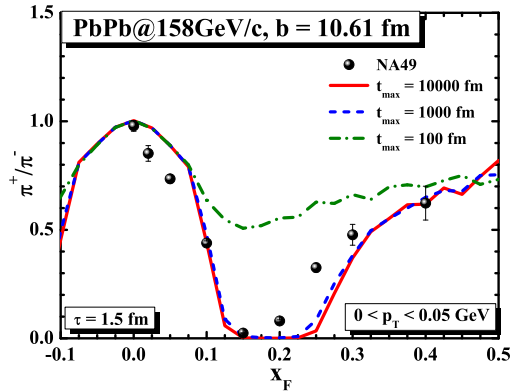}
\vspace{-1.9cm} \caption{ The $\pi^+ / \pi^-$ ratio as a function of
the time of the evolution $t_{max}$. Here $\tau$ = {1.5 fm/$c$ was used
for an example}. } \label{fig:ratio_tmax}
%(see Sec.~\ref{long} for comparison}). } \label{fig:ratio_tmax}
\end{figure}
%-----------------------------------------------------------------------

In the center-of-mass system one needs to include both electric and
magnetic fields generated by the moving
spectators~\cite{EM_previous_1}. In the present calculation we
assume that spectators move with the velocity of their initial
parent beams. We perform the evolution of charged pions in the
electromagnetic field generated by spectators separately for
positively and negatively charged pions. After the evolution of pion
trajectory is terminated a final event is generated. The events are
used then to generate momentum spectra of $\pi^+$ and $\pi^-$ in the
final state of the collision.

To summarize, we generate 100 million pions using a weighted Monte
Carlo code with the corresponding weight:
\begin{equation}
W=\frac{dN}{dy}\cdot\frac{dN}{dp_T}^{weighted}, \label{weight}
\end{equation}
where $dN/dy$ is given by Eqs.~(\ref{fragmentation_function}) and
(\ref{pion_rapidity}). To be consistent with the fire-streak model
one has to keep the normalization of the weighted
transverse-momentum spectrum of pions equal to unity, i.e.
\begin{equation}
\frac{dN}{dp_T}^{weighted}\equiv\frac{1}{S}\frac{dN}{dp_T},
\end{equation}
where $dN/dp_T$ is given by Eq.~(\ref{fit_function}) and its
normalization is defined in Eq.~(\ref{normalization}). We generate
the rapidity and transverse momentum of pions using an uniform
random number generator in the following ranges of kinematical
parameters: $y\in[-5;5]$, $p_T\in[0;1.1]$~GeV/$c$,
$\phi\in[0,2\pi]$.

%-----------------------
\section{Results and discussion}
\label{results}
%-----------------------

%-------------------------------------------------
\subsection{Simplified static source}
%-------------------------------------------------
\label{static}

%-------------------------------------------------------------------
\begin{figure}
\includegraphics[width=0.5\textwidth]{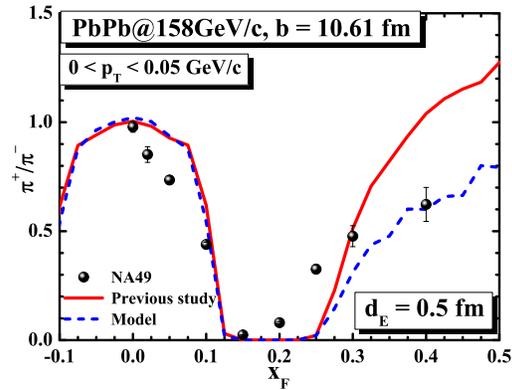}
\vspace{-1.9cm} \caption{The $\pi^+/\pi^-$ ratio as a function of
$x_F$ for $0<p_T<0.05$~GeV/$c$ obtained in our previous
study~\cite{EM_previous_1} (solid red line) as well as from the
%partially simplified
{intermediate}
model {scenario} described in the text (dashed blue line).
Both simulations are taken with the distance $d_E$ set to 0.5~fm.
Simulation results are put in comparison to experimental data from
NA49 (symbols).}\label{de05_comparison}
%\vspace*{-0.5cm}
\end{figure}
%-------------------------------------------------------------------
%------------------------------------------------------------------------
\begin{figure}
%[b!]
\includegraphics[width=0.5\textwidth]{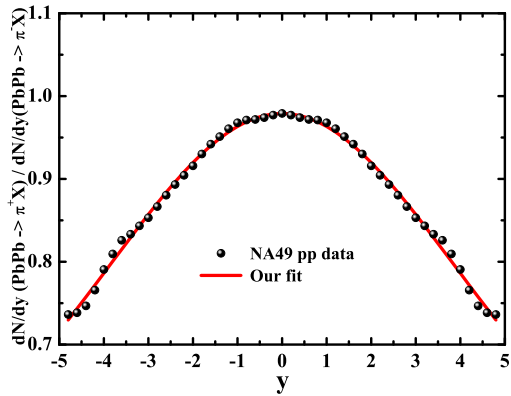}
\vspace{-1.9cm}\caption{The ratio~(\ref{isospin_a}) as a function of
pion rapidity. The black symbols correspond to the results obtained
by using NA49 $pp$ data~\cite{NA49_pp} (the corresponding procedure
is described in the text), the solid red line represents our fit to
the pseudodata.} \label{nuclear_ratio}
\end{figure}
%-----------------------------------------------------------------------
%-----------------------------------------------------------------------

\begin{figure*}
%\begin{figure}[b!]
\vspace{0.3cm} \centering \subfigure{
\resizebox{0.47\textwidth}{!}{%
 \includegraphics{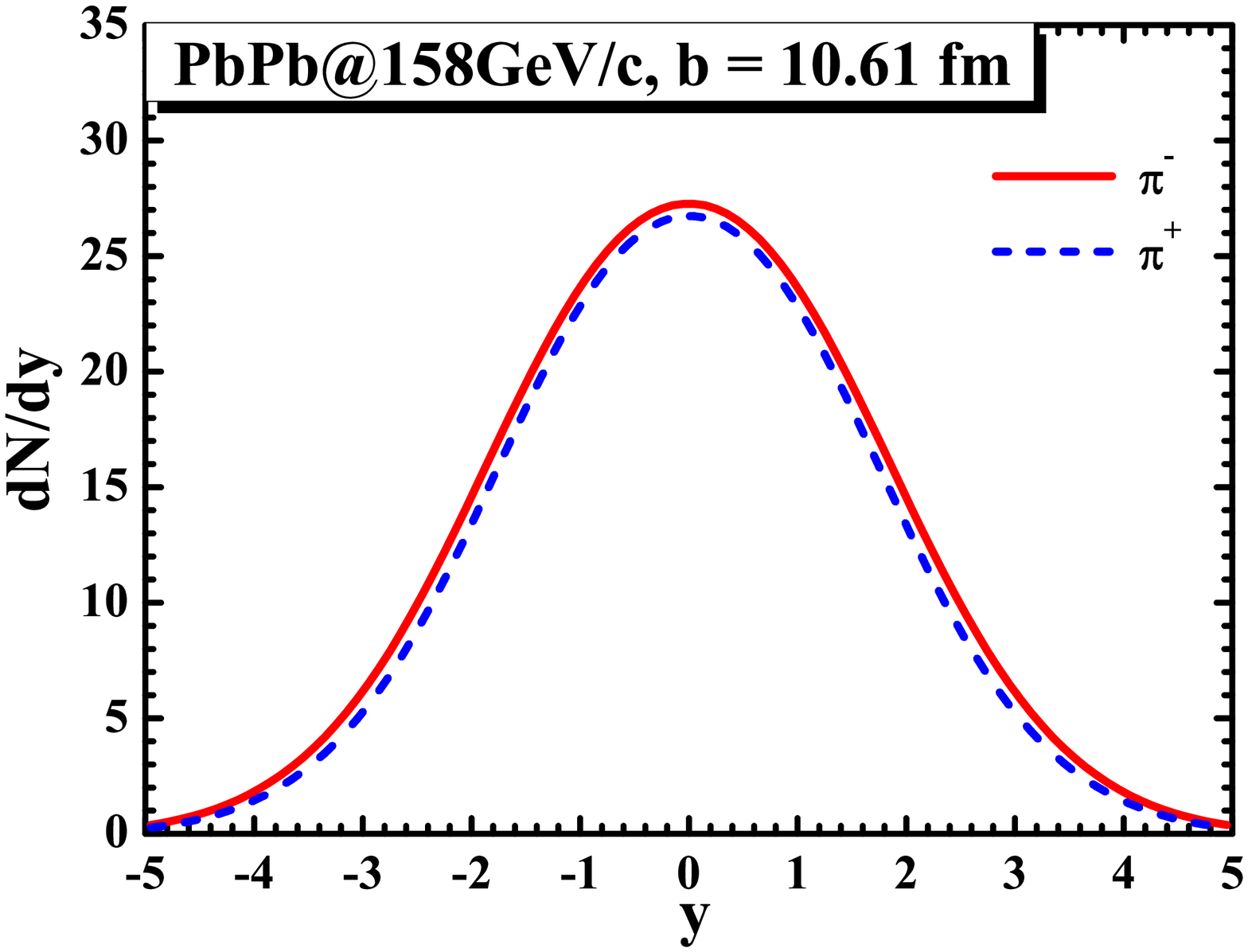}\hspace*{-5cm}
} } \subfigure{
\resizebox{0.47\textwidth}{!}{%
 \includegraphics{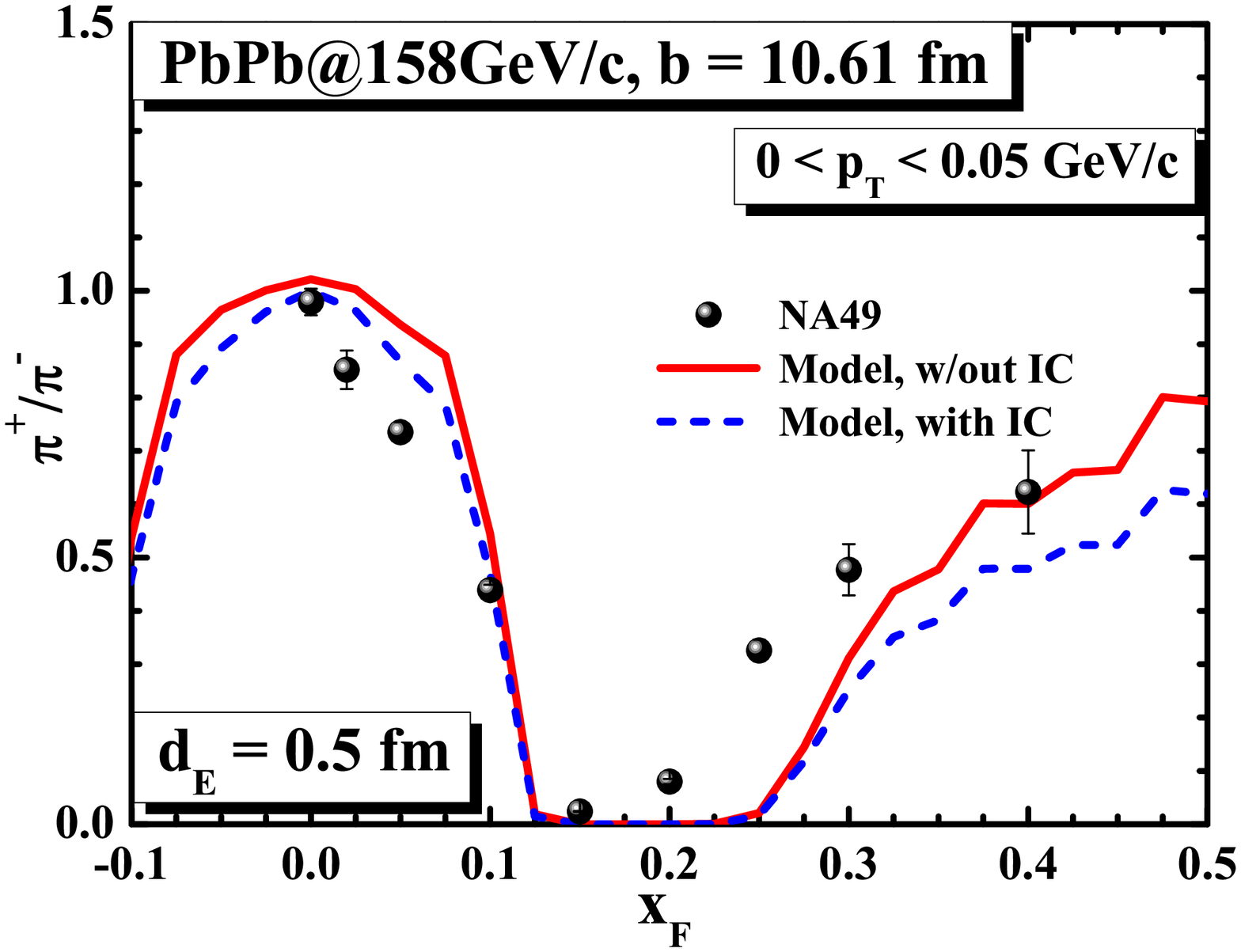}\hspace*{-5cm}
} } \vspace{-1.7cm}\caption{Left: the isospin-corrected rapidity
distribution of $\pi^+$ (dashed blue line) in comparison to the
rapidity distribution of $\pi^-$ (solid red line) as a function of
pion rapidity, obtained from the fire-streak model as described in
the text. Right: the final-state $\pi^+/\pi^-$ ratio as a function
of $x_F$ for $0<p_T<0.05$~GeV/$c$, calculated without (solid red
line) and with (dashed blue line) isospin corrections.}
\label{isospin}
\end{figure*}
%
%-----------------------------------------------------------------------

We start with the comparison of our present work to the previous
results published in Ref.~\cite{EM_previous_1}. In the previous
study the pion emission source was reduced to a single point in
space: $(x,y,z)=(0,0,0)$. The initial two-dimensional $(x_F,p_T)$
distribution of pions was assumed similar to that in nucleon-nucleon
collisions (the same distribution was assumed for $\pi^+$ and
$\pi^-$). The time of pion emission, directly equivalent to the
distance $d_E$ between the pion source and the two spectator
systems, was taken as a free parameter. The result for the
$\pi^+/\pi^-$ ratio as a function of $x_F$ for the range of
transverse momentum of pions, $0<p_T<0.05$~GeV/$c$, for this
scenario and assuming $d_E=0.5$~fm is presented in
Fig.~\ref{de05_comparison} (solid red line), in comparison to the
NA49 experimental data~\cite{NA49_EM}.

In order to illustrate the role of the different elements of our
fire-streak model in the description of the electromagnetic effect
on $\pi^+/\pi^-$ ratios, our comparison with the previous study will
be made in a few successive steps. In the first step, we consider a
specific,
%partially simplified scenario
{\em intermediate} model
scenario where the pion source is
extended in the transverse direction, but {\em static}, and {fixed at
$z=0$}. The transverse extent of our source is defined by the initial
transverse positions of the fire streaks in the $(x,y)$ plane, given
by the overlap of the two nuclei (see Ref.~\cite{FS_model_1} for a
detailed discussion). The pion emission time is arbitrarily set to
0.5~fm/$c$ which very closely corresponds to $d_E=0.5$~fm. Finally,
the initial two-dimensional $(y,p_T)$ distribution of pions is
initialized according to Eq.~(\ref{weight}) with the rapidity
distribution taken from the complete fire-streak model,
Eq.~(\ref{pion_rapidity}), and with the transverse-momentum distribution
taken from the UrQMD model.
We underline the {\em intermediate} character of this step of our
analysis as for the time being we keep the static source used
in the previous study~\cite{EM_previous_1}; this results in
a single value of $d_E$. This is to be opposed to
the full fire-streak model scenario which we will discuss later,
in Sec.~\ref{long}.

The results for the $\pi^+/\pi^-$ ratio
as a function of $x_F$ are shown in Fig.~\ref{de05_comparison} as
the dashed blue line. It is clear from the figure that the inclusion
of the transverse extent of the source and a more realistic
description of the initial rapidity and transverse momentum
distributions has a visible effect on the predictions of the model
above $x_F=0.25$. In this context we note that these results were
obtained assuming the initial rapidity distribution for positive
pions equal to that of negative pions.

\vspace*{-0.4cm}
%----------------------------------------
\subsection{Isospin correction}
%----------------------------------------

Consequently, as a next step we have to implement the {\em isospin
correction}, namely the difference between rapidity distributions of
positive and negative pions initially produced from the fire
streaks. We proceed as follows.
First we note that generally in $p+p$ collisions at SPS energies:
\begin{equation}
\frac{d \sigma_{p p \to \pi^+X}}{d y d p_T} \ne \frac{d \sigma_{p p
\to \pi^-X}}{d y d p_T} \; . \label{asymmetry}
\end{equation}
This fact is well known experimentally~\cite{NA49_pp}, but can be
also explained by QCD-based calculations~\cite{czech}. Therefore,
also for two colliding $Pb$ nuclei ($Z=82$, $A-Z=126$) some
discrepancy between $\pi^+$ and $\pi^-$ distributions is to be
considered, and seems indeed indicated by the compilation of
numerical results from the NA49 experiment~\cite{num49}.
As no rapidity distribution of $\pi^+$ was measured for $Pb+Pb$
collisions at top SPS energy, we assume that the latter can be
approximated by postulating the $y$-dependence of the $\pi^+/\pi^-$
ratio in $Pb+Pb$ reactions to be similar to that in the proper
superposition of $p+p$, $n+p$, $p+n$ and $n+n$ collisions.
We underline that the above assumption is made only for the
$\pi^+/\pi^-$ ratio rather than for $\pi^+$ and $\pi^-$ yields.
Following the approach proposed in Refs~\cite{isospin,FS_model_2},
and invoking isospin symmetry in pion production for participating
protons and neutrons $(n\rightarrow\pi^-=p\rightarrow\pi^+)$, this
ratio reads:
\begin{eqnarray}
\frac{\frac{dN}{dy}(PbPb\rightarrow\pi^+X)}{\frac{dN}{dy}(PbPb\rightarrow\pi^-X)}
=\hspace*{4cm}\vspace*{1.3cm}\nonumber\\
\frac{Z\frac{dN}{dy}(pp\rightarrow\pi^+X)+(A-Z)\frac{dN}{dy}(pp\rightarrow\pi^-X)}{Z\frac{dN}{dy}(pp\rightarrow\pi^-X)+(A-Z)\frac{dN}{dy}(pp\rightarrow\pi^+X)},
\label{isospin_a}
\end{eqnarray}
where $Z=82$ and $A=208$ for the considered case of $Pb+Pb$
collisions. We note that the above formula~(\ref{isospin_a}) is,
within experimental uncertainties, valid for ratios of total $\pi^+$
over $\pi^-$ multiplicities
%, and $\pi^+$ over $\pi^-$ mid-rapidity yields
in peripheral $Pb+Pb$ reactions provided by the NA49
Collaboration~\cite{num49}. It is also equivalent to the prediction
of the Wounded Nucleon Model~\cite{wnm} for $\pi^+/\pi^-$ ratios as
a function of rapidity, once the isospin differences between protons
and neutrons are included in this model.

Having the NA49 experimental data for rapidity distributions of
positive and negative pions for the proton-proton
collisions~\cite{NA49_pp} and inserting them to
Eq.~(\ref{isospin_a}), we construct the assumed $\pi^+/\pi^-$ ratio
in peripheral $Pb+Pb$ reactions.
This we show in Fig.~\ref{nuclear_ratio} as a function of rapidity.
The $\pi^+/\pi^-$ ratio in  $Pb+Pb$ collisions obtained
from the NA49 $p+p$ data is shown as the black symbols.
A purely mathematical fit to the pseudodata (solid red line)
will be used to parametrize this ratio.
We notice a rather sizeable $\pi^+$-$\pi^-$ asymmetry at large
rapidity.

Taking into account the fit shown in Fig.~\ref{nuclear_ratio}, we
can construct the $\pi^+$ rapidity distribution using the
fire-streak model addressed in section~\ref{Rap-distr-pions}. We
postulate that in analogy to the fragmentation function for negative
pions, our isospin-corrected fire-streak fragmentation function into
positive pions can be written as:
\begin{equation}
\frac{dn}{dy}(y,y_s,E_s^*,m_s\!)\!=\!A(E_s^*-m_s\!)\exp\biggl(\!\!-I_c\frac{[(y\!-\!y_s)^2\!+\!\epsilon^2]^{\frac{r}{2}}}{r\sigma_y^r}\!\biggr),
\label{fragmentation_function_isospin}
\end{equation}
where $I_c$ will be treated as a new adjustable parameter. Just as a
reminder, for negative pions $I_c=1$ was assumed in
Eq.~(\ref{fragmentation_function}) by definition. We find that once
$I_c=1.175$ is taken for positive pions, our fire-streak model
provides a good (within 2-3\%) description of the fit shown in
Fig.~\ref{nuclear_ratio}.

In Fig.~\ref{isospin} (left plot) we present the resulting
comparison of the two initial rapidity distributions: the
isospin-corrected distribution of positive pions (dashed blue line)
and the distribution of negative pions (solid red line), both
calculated within the fire-streak model as described above.

The effect of the isospin correction on the result of our simulation
is demonstrated in Fig.~\ref{isospin} (right plot), where we show
the resulting final state $\pi^+/\pi^-$ ratio without (solid red
line) and with (dashed blue line) isospin correction as a function
of $x_F$. For both calculations the same
simplified
%{intermediate}
static source
model is used as described in section~\ref{static}. We note that our
isospin correction on the initial rapidity spectrum of $\pi^+$ gives
quite a sizeable effect on the calculated $\pi^+/\pi^-$ ratio, after
the inclusion of the electromagnetic effect caused by spectators.

%-------------------------------------------------------------------
\subsection{Longitudinal evolution of the system}
%-------------------------------------------------------------------
\label{long}

All the results presented so far were simulated with a partially
simplified, intermediate model where the pion emission source was
static in position space, and fixed at $z=0$. The pion emission time/distance
was set to $d_E=0.5$~fm in the collision center-of-mass system. Now
we finally include the longitudinal evolution of the system along
the $z$ axis, given by the fire-streak velocities
(Fig.~\ref{schematic_collision}). Consequently we have to introduce
the pion creation time, $\tau$, which we define for each fire streak
in its own rest frame. First we perform the simulations for various
fixed values of $\tau$: $\tau=0.5; 1; 1.5; 2$~fm/$c$ (please note
that this way we assume the pion creation time is the same in the
rest frame of each fire streak). The results of the simulation are
presented in Fig.~\ref{tau_comparison}.
As now they correspond to the complete fire-streak model
(in its formulation which we proposed in
Ref.~\cite{FS_model_1}),
we label them ``FS'' to differentiate from the
simplified intermediate scenario we used previously.

In comparison to Fig.~\ref{isospin} (right plot), we state that the
inclusion of the longitudinal evolution of the system influences the
observed electromagnetic distortion, and tends to increase the
$\pi^+/\pi^-$ ratios for pions at high $x_F$
%pions
in the final state. On the other hand, after a detailed inspection
of Fig.~\ref{tau_comparison}, one can conclude that there is no
configuration with fixed pion creation time that can well describe
the experimental data.

%------------------------------------------------------------------
\begin{figure}
%\vspace{0.5cm}
\includegraphics[width=0.5\textwidth]{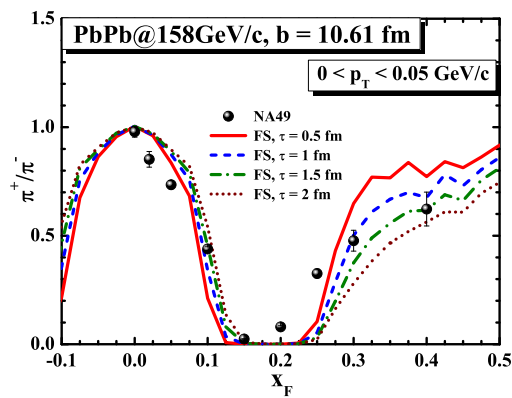}
\vspace{-1.9cm} \caption{The $\pi^+/\pi^-$ ratio as a function of
$x_F$ for $0<p_T<0.05$~GeV/$c$ calculated within
our version of
the fire-streak
(FS) model with different pion creation times, in comparison to the
NA49 data~\cite{NA49_EM}.}\label{tau_comparison}
\end{figure}
%------------------------------------------------------------------

%-------------------------------------------------------------
\begin{figure*}
\vspace{0.3cm} \centering \subfigure{
\resizebox{0.47\textwidth}{!}{%
 \includegraphics{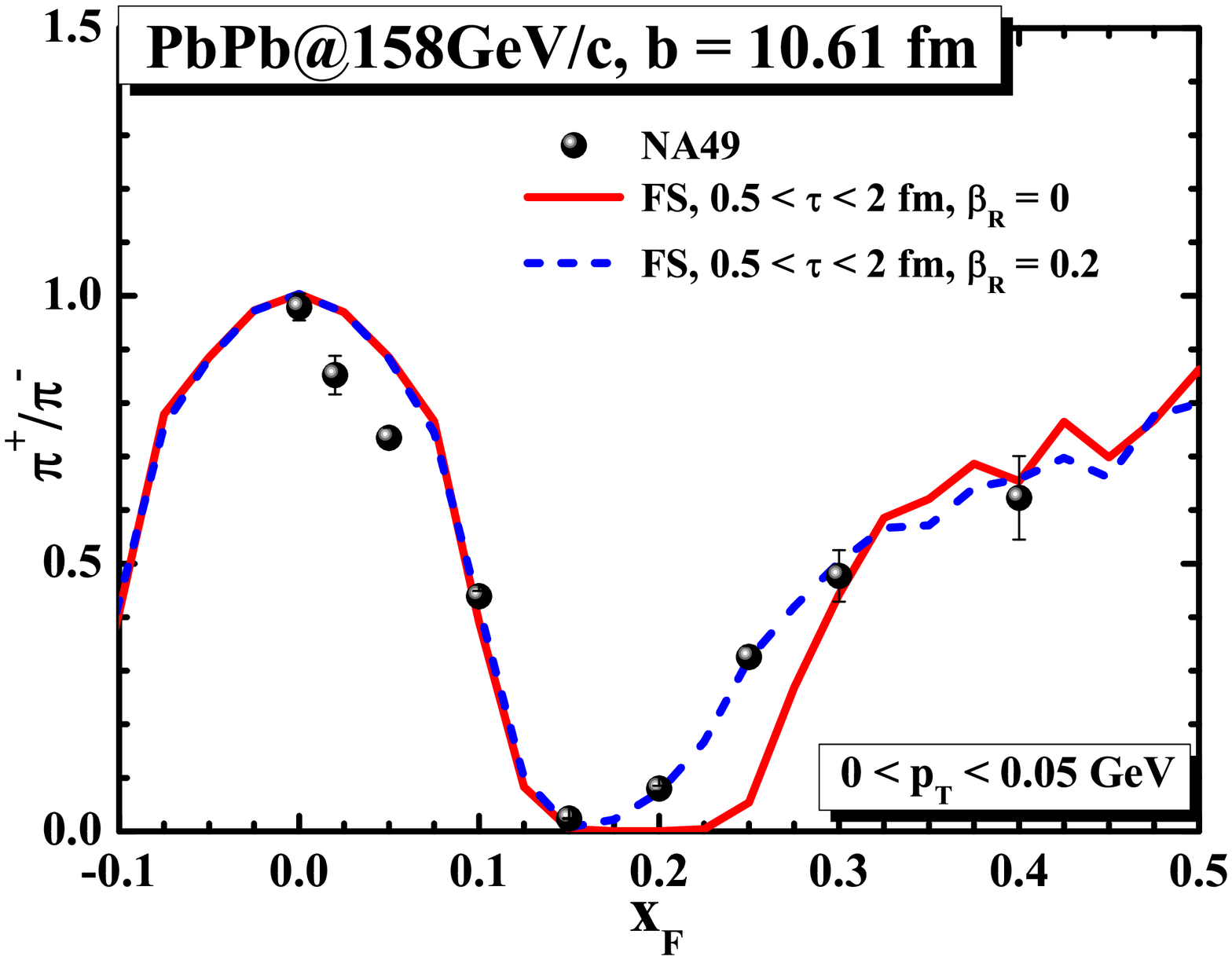}\hspace*{-5cm}
} } \subfigure{
\resizebox{0.47\textwidth}{!}{%
 \includegraphics{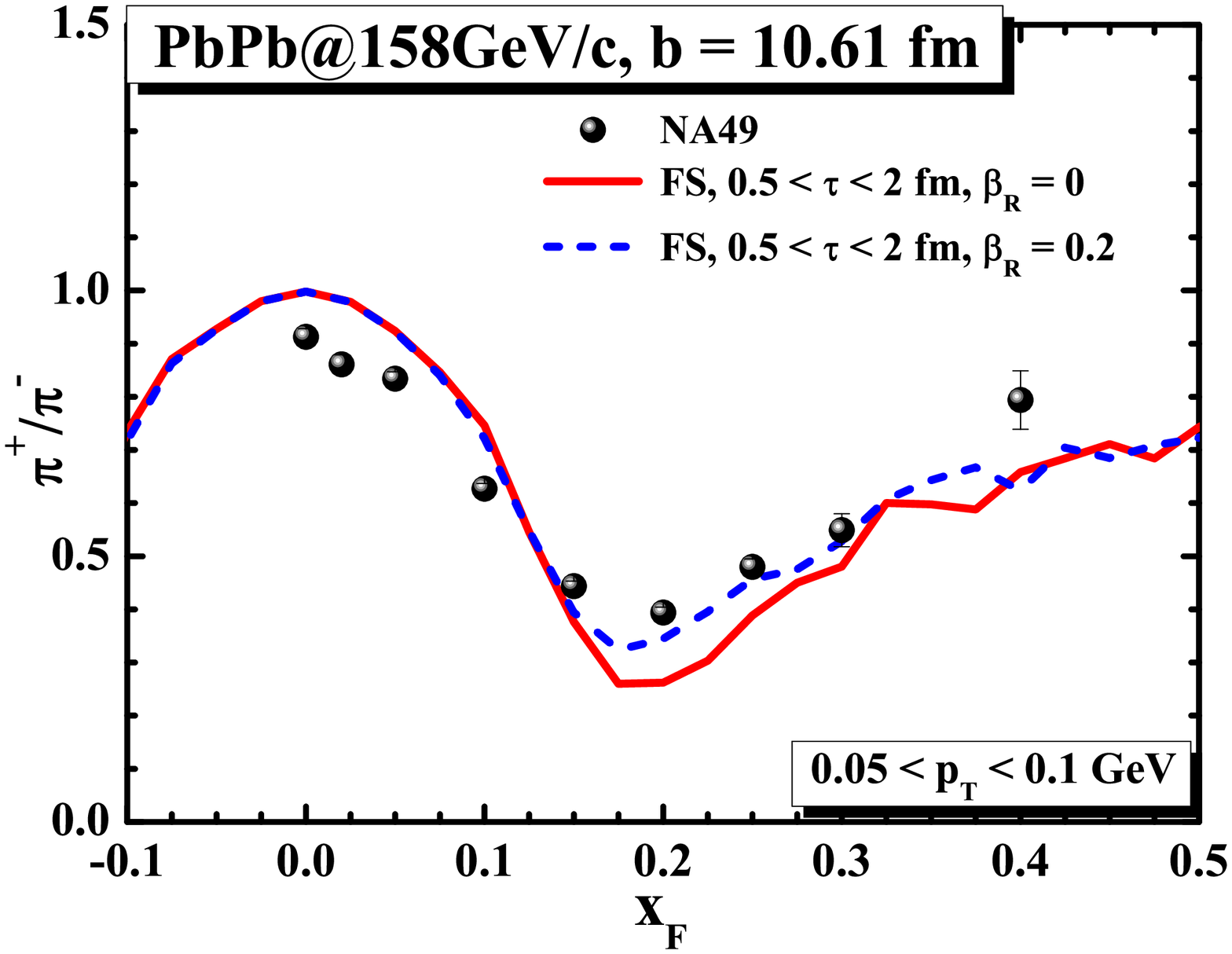}\hspace*{-5cm}
} } \vspace{-1.7cm}\caption{The results of calculation of
electromagnetic effects on the $\pi^+/\pi^-$ ratio as a function of
$x_F$ in peripheral $Pb+Pb$ collisions at top SPS energy, obtained
within
our version of
the fire-streak (FS) model with the pion creation time given
by Eq.~(\ref{tau_dependence}) for two different ranges of pion
transverse momentum: $0<p_T<0.05$~GeV/$c$ (left plot) and
$0.05<p_T<0.1$~GeV/$c$ (right plot). The scenario with stable
spectators is shown by the solid red lines, and the scenario with
expanding spectators with the radial surface velocity,
$\beta_R=0.2$, is presented by the dashed blue lines.} \label{final}
\end{figure*}
%-------------------------------------------------------------
%\newpage

It seems indeed difficult to expect that all the longitudinal
elements of excited matter will be characterized by the same pion
creation time $\tau$. In fact we expect that the fire-streak
lifetime increases with its excitation energy, that is
\begin{equation}
\tau = F(E_s^*-m_s),
\end{equation}
where $F$ is a monotonically increasing function and $E_s^*$ and
$m_s$ stand respectively for the total energy and the ``cold mass''
of the fire streak, as described in
Eq.~(\ref{fragmentation_function}).
For this reason, we try to simulate an initial configuration with
the pion creation time which is not fixed. For the present work, we
choose the following simple linear dependence:
\begin{equation}
\tau=a(E_s^*-m_s)+\tau_0, \label{tau_dependence}
\end{equation}
where $\tau_0=\tau_{min}=0.5$~fm/$c$ and $\tau_{max}$ is set to be
$2$~fm$/c$, which gives us $a\approx0.08$~(for energies given in
GeV).

In Fig.~\ref{final} we present (by the solid red lines) the results
of calculation of the electromagnetic effect on the $\pi^+/\pi^-$
ratio as a function of $x_F$ in peripheral $Pb+Pb$ collisions at top
SPS energies, with the pion creation time parametrized as in
Eq.~(\ref{tau_dependence}). The results are shown for two different
ranges of pion transverse momentum: $0<p_T<0.05$~GeV/$c$ (left plot)
and $0.05<p_T<0.1$~GeV/$c$ (right plot).
While generally, the solid red line reproduces the main features of
the $x_F$- and $p_T$-dependence of the electromagnetic effect, the
detailed shape of the minimum at $x_F\approx 0.15-0.2$ is still
rather poorly described by the simulation.
At this point we notice that in the calculations made so far we
always assumed that spectators were {\em stable}, at least on the
time scale when electromagnetic fields interact with charged pions
(in our simulation the spectator systems were taken as two stable,
homogeneously charged spheres as described in
Ref.~\cite{EM_previous_1}). However, the spectators are rather
highly excited systems~\cite{Mazurek}. Therefore, presently we
impose a scenario with expansion of the spectators.
In its own rest frame, each of the two spectator systems is taken as
a homogeneously charged sphere, expanding radially with a given
surface velocity $\beta_R$.
We introduce this surface velocity $\beta_R$ as an additional free
parameter. The configuration with $\tau$ given by
Eq.~(\ref{tau_dependence}) and $\beta_R=0.2$ (in the spectator rest
frame) gives the best description of the NA49 experimental
data~\cite{NA49_EM} for both ranges of pion transverse momentum,
$0<p_T<0.05$~GeV/$c$ and $0.05<p_T<0.1$~GeV/$c$, as shown in
Fig.~\ref{final} by the dashed blue lines.
We note that the surface velocity $\beta_R=0.2$ corresponds, for the
radially expanding sphere, to a mean velocity of spectator expansion
of 0.15$c$. This is reminiscent of the characteristic spectator
expansion velocity of 0.16$c$ obtained by Cugnon and Koonin at a
much lower collision energy~\cite{cugnon81}.

%{\it {\rm Finally,} in Fig.\ref{fig:ratio_flow} we demonstrate effect of the pion flow on the observed $\pi^+ / \pi^-$ ratio as a function of Feynman $x_F$, separately for $v_2$ and $v_1$. We observe no effect for $v_2$ (dash-dotted line) and nonnegligible effect for $v_1$ (dashed line). The reason for this is the following. As shown in Fig.\ref{fig:ratio_flow} $v_1$ is large at large rapidities and diverges quickly from $v_1$ = 0 at $p_t$ = 0. Vice versa $v_2$ is small at large rapidities $y$ and grows rather slowly with $p_t$, being 0 at $p_t$ = 0.

Finally, in Fig.~\ref{fig:ratio_flow} we demonstrate
the effect of the pion flow on the observed $\pi^+ / \pi^-$ ratio
as a function of Feynman $x_F$, for $v_1$ and $v_2$ included
together. We find no effect for $v_2$, and {a non-negligible}
effect for $v_1$. The reason for this is the following. As shown in
Fig.~\ref{fig:flow_fit}, $v_1$ is large at large rapidities and
diverges quickly from $v_1$ = 0 at $p_T$ = 0. Vice versa $v_2$ is
small at large rapidities $y$ and grows rather slowly with $p_T$,
being 0 at $p_T$ = 0.

%----------------------------------------------------------------------
\begin{figure*}
\vspace{0.3cm} \centering \subfigure{
\resizebox{0.47\textwidth}{!}{%
%resizebox{0.47\textwidth}{!}{%
%\includegraphics{pion_ratio_final_flow_pt_0_025GeV.eps}\hspace*{-5cm}
 \includegraphics{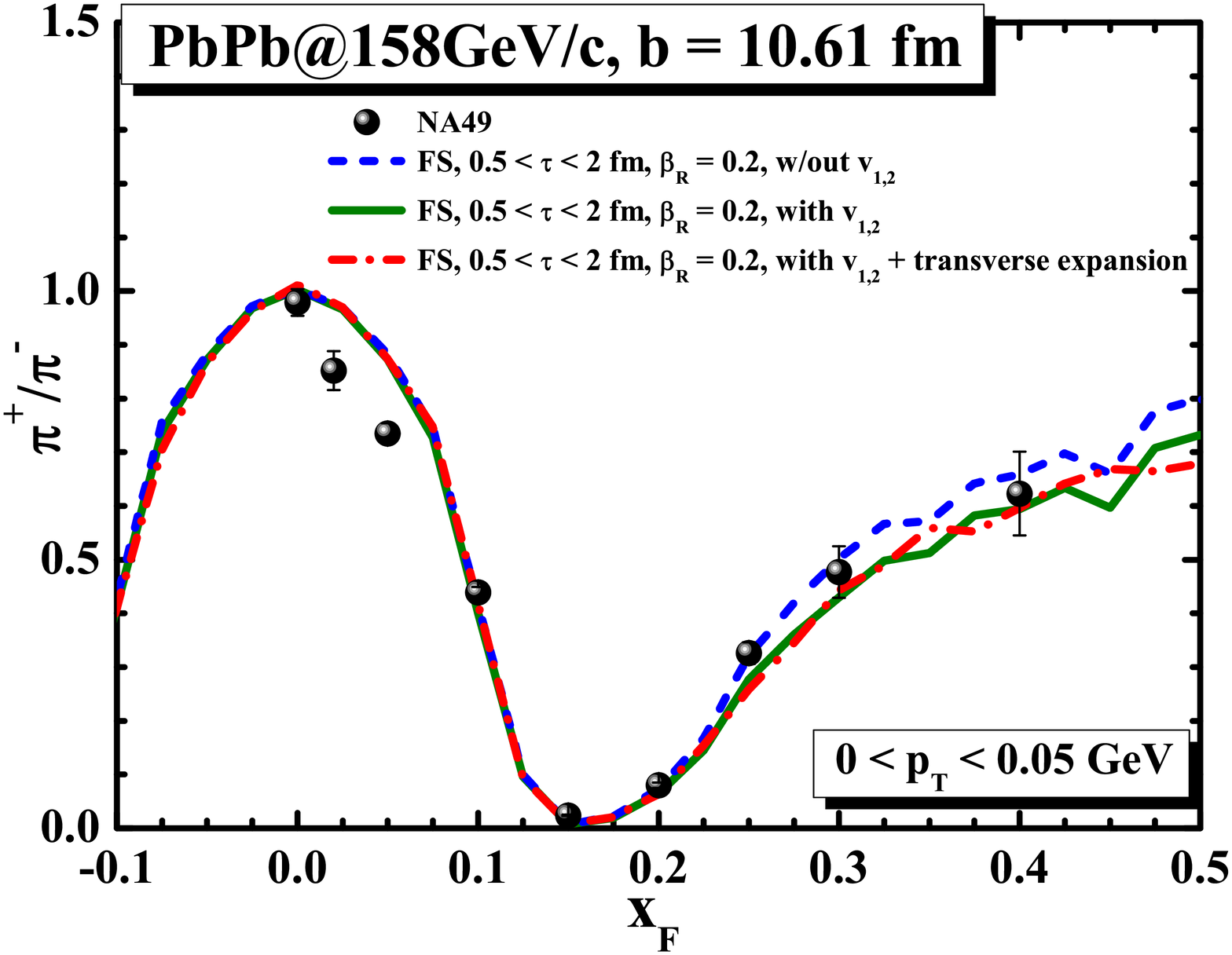}\hspace*{1cm}
} } \subfigure{
\resizebox{0.47\textwidth}{!}{%
%\resizebox{0.47\textwidth}{!}{%
%\includegraphics{pion_ratio_final_flow_pt_0_075GeV.eps}\hspace*{-5cm}
 \includegraphics{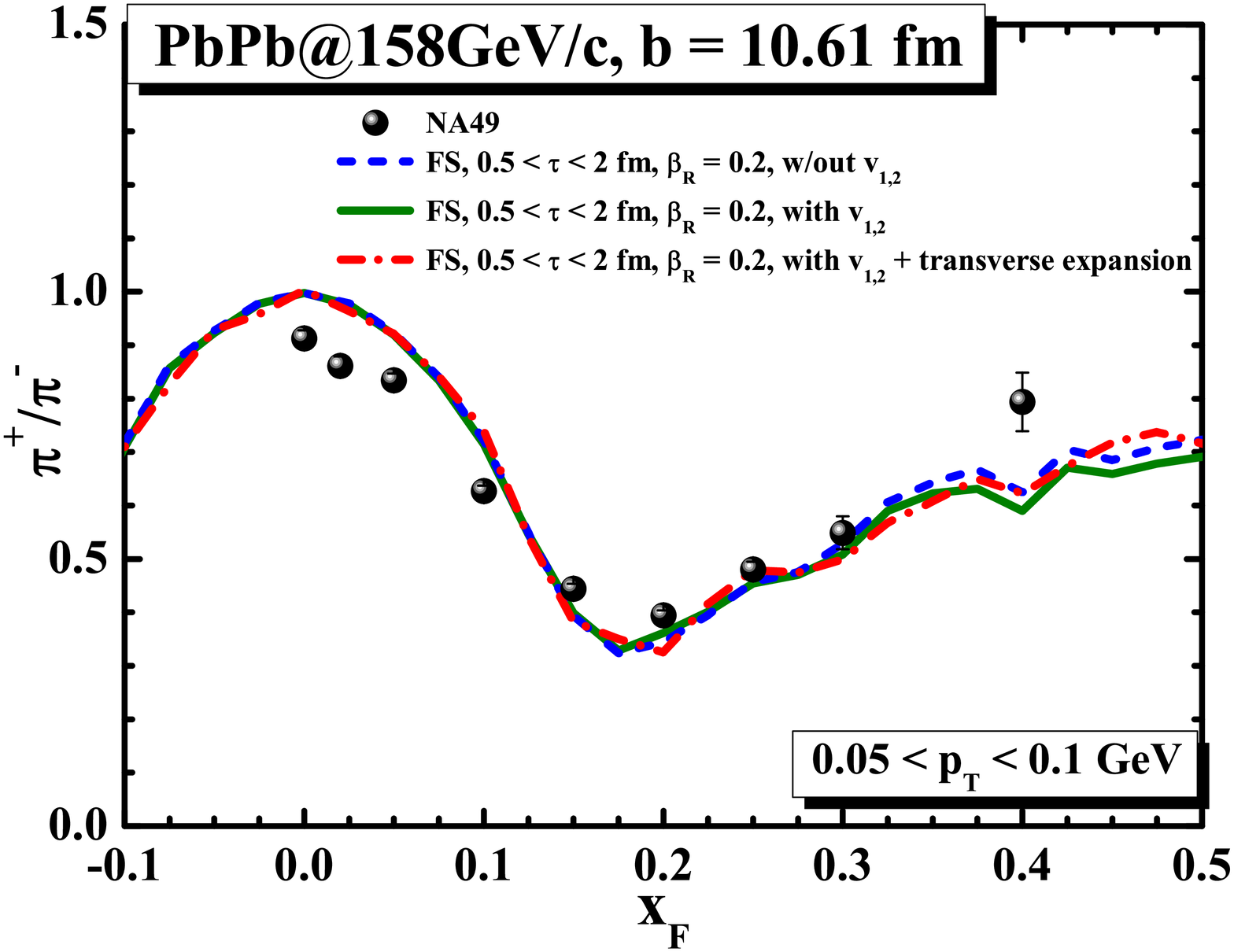}\hspace*{1cm}
} } \vspace{-0.0cm}\caption{The effect of inclusion of the initial
$v_1$ and $v_2$ flow coefficients on the observed $\pi^+/\pi^-$
ratio (solid green lines). The dashed blue lines are for the case
when no flow is taken into account (the same as in
Fig.~\ref{final}). The dash-dotted red lines include the effect of
change of the initial position of pions due to transverse expansion
of the fire streaks. } \label{fig:ratio_flow}
\end{figure*}
%-----------------------------------------------------------------------

The present description of the experimental data by our simulation
is satisfactory for faster pions (with $x_F\geq 0.1$). Our present
conclusion is that in this kinematic region, the electromagnetic
distortion of $\pi^+/\pi^-$ ratios can be successfully described
once five basic components are taken into account: (1) a realistic
description of the longitudinal evolution of the system provided by
the fire-streak model, (2) isospin differences between initial
$\pi^+$ and $\pi^-$ emission, (3) a proper pion creation time, (4)
the expansion of the spectator system, and (5) charged pion
propagation through the electromagnetic field until relatively long
times ($\sim10^4$~fm/$c$). We note that relatively short pion
creation times ($0.5<\tau<2$~fm/$c$) are needed the explain the
experimental data. This is in contrast with significantly longer
decoupling times obtained from other methods~\cite{hbt}. We
interpret this difference as due to the fact that unlike for the
cited papers, our study is in practice anchored to the regime of low
pion transverse momenta and high rapidities ($x_F\geq 0.1$,
$y\gtrsim y_{beam}$), most sensitive to the spectator charge.
Finally, we remark that our simulation noticeably overpredicts the
measured $\pi^+/\pi^-$ ratios in the region $x_F\leq 0.05$, that is
at central rapidities. This we attribute to the effect of
participant charge, reported before at SPS energies~\cite{na44}.
We leave this effect for a future analysis.

%-------------------------------------------------------------------
\subsection{Transverse expansion}
%-------------------------------------------------------------------
\label{trexp}

In the present and following section we discuss the possible
influence of two additional effects on the results of our
calculation. Firstly we address the issue of the transverse
expansion of the fire-streak matter, up to now neglected in our
analysis. This we estimate by modifying the transverse position of
pion emission points at the pion emission time $\tau$, with respect
to the original fire-streak transverse position $(b_x,b_y)$.
Consequently, the transverse coordinates of the pion emission point
are taken as:
\begin{equation}
x=b_x+\Delta x ,
\end{equation}
\vspace*{-1cm}
\begin{equation}
y=b_y+\Delta y .
\label{trexp1}
\end{equation}
Where the displacements $\Delta x$ and $\Delta y$ are randomly generated from a flat two-dimensional distribution within a maximum radius $R$ given as
\begin{equation}
R = r_0 + \beta_{tr}\cdot\tau .
\label{trexp2}
\end{equation}
where we set $r_0=0.5$~fm as representative for the original
transverse size of the fire streak while $\beta_{tr}$ is tentatively
set to 0.5$c$. The result of this modification is shown as
dash-dotted line in Fig.~\ref{fig:ratio_flow}. The corresponding
effect appears basically negligible, which we interpret as a
consequence of the fact that the transverse expansion of the fire
streak as modelled above will not modify the average transverse
position of the pion emission point with respect to the spectator
system, while the broadening of the corresponding distribution of
$(x,y)$ points will remain moderate for the values of $\tau$
discussed in section~\ref{long}. Thus we conclude that transverse
expansion of fire streaks will not modify our conclusions
significantly.

%-------------------------------------------------------------------
\subsection{Vorticity of fire streaks}
%-------------------------------------------------------------------
\label{vort}

Another effect to be considered is the well-known phenomenon of
vorticity of strongly-interacting matter created in the
collision~\cite{star-nature}. This effect is potentially important
as the rotation of fire streaks driven by shear viscosity, even for
a short time in the early stage of the collision, will modify their
trajectories and consequently also the position of the pion emission
points with respect to the spectator system.

%------------------------------------------------------------------
\begin{figure}
%\vspace{0.5cm}
\includegraphics[width=0.4\textwidth]{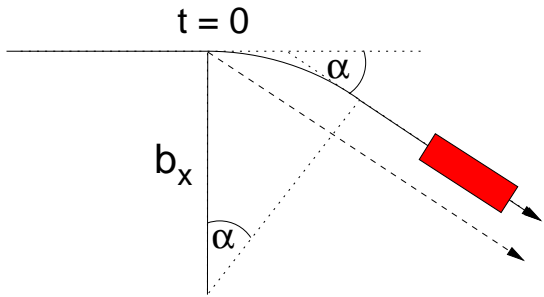}
\vspace{-0.0cm} \caption{Schematic drawing of the modification of fire-streak trajectory by rotation around the geometric center of the collision.}\label{drawing-r}
\end{figure}
%------------------------------------------------------------------

%------------------------------------------------------------------
\begin{figure}
%\vspace{0.5cm}
%\includegraphics[width=0.48\textwidth]{v1_y_rotation_2.pdf}
\includegraphics[width=0.48\textwidth]{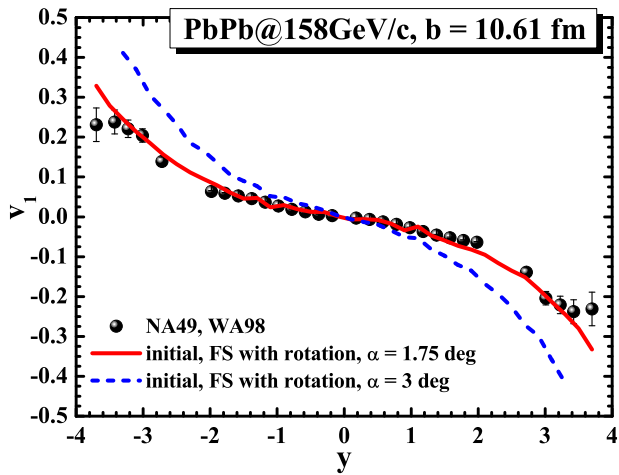}
\vspace{-0.2cm} \caption{Directed flow of initially emitted pions as a function of rapidity obtained by our simulation assuming fire-streak rotation by $\alpha=1.75\deg$,
%compared to
{superimposed with}
the experimental data from Refs.~\cite{NA49_flow,WA98_flow}.
Our prediction for $\alpha=3\deg$ is also
%included
shown
for comparison.}\label{v1_y_rotation}
\end{figure}
%------------------------------------------------------------------

%----------------------------------------------------------------------
\begin{figure*}
\vspace{0.3cm} \centering \subfigure{
\resizebox{0.47\textwidth}{!}{%
%resizebox{0.47\textwidth}{!}{%
%\includegraphics{pion_ratio_final_flow_pt_0_025GeV.eps}\hspace*{-5cm}
 \includegraphics{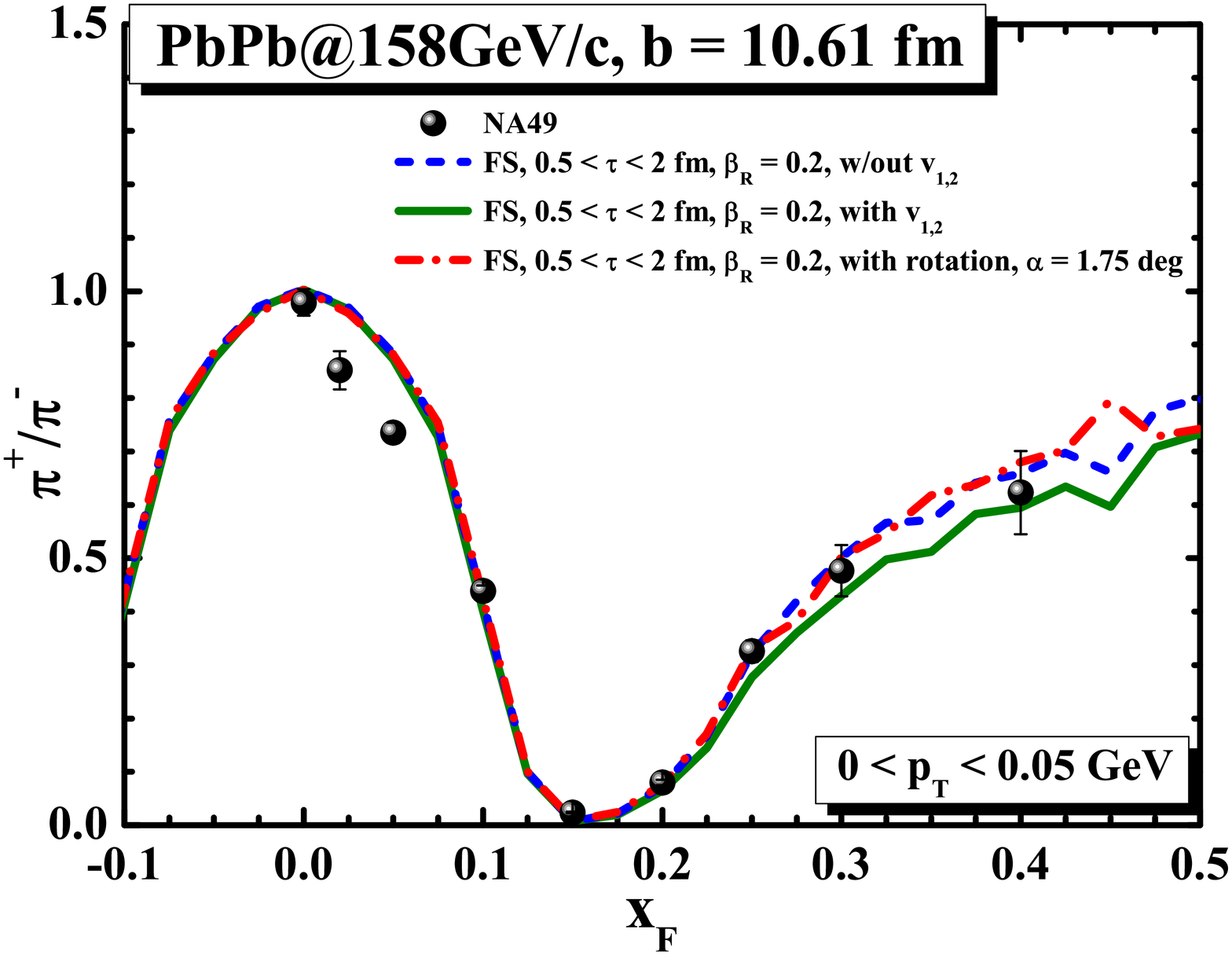}\hspace*{1cm}
} } \subfigure{
\resizebox{0.47\textwidth}{!}{%
%\resizebox{0.47\textwidth}{!}{%
%\includegraphics{pion_ratio_final_flow_pt_0_075GeV.eps}\hspace*{-5cm}
 \includegraphics{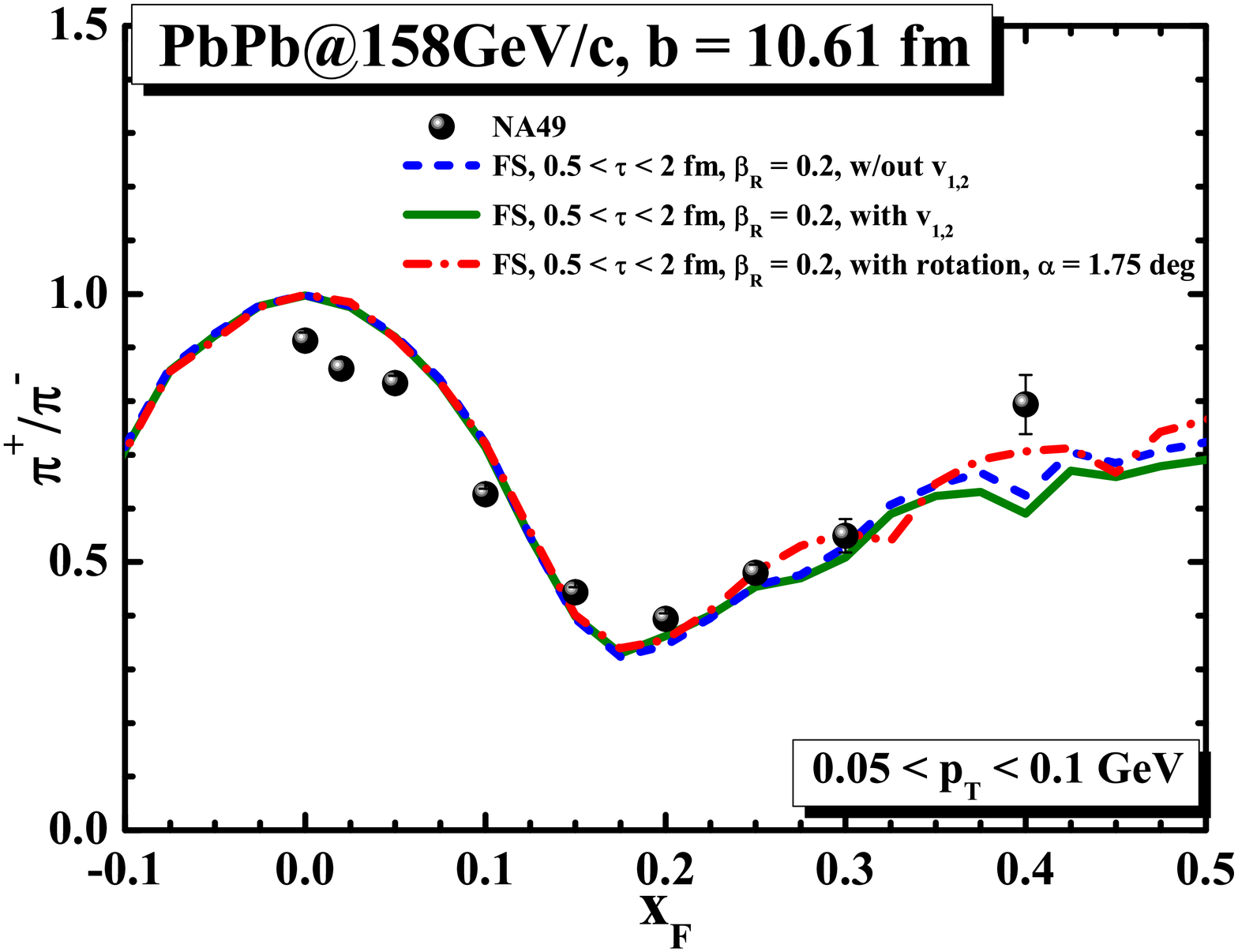}\hspace*{1cm}
} } \vspace{-0.0cm}\caption{The effect of inclusion of fire-streak
rotation as described in the text on the observed $\pi^+/\pi^-$ ratio
(dash-dotted red lines).
The dashed blue lines are for the case when no flow is taken into account and the
%(the same as in Fig.~\ref{final}).
%The
green solid lines are for the case when flow is taken into account in the way
described in Sec.~\ref{long} (both dashed blue and green solid lines are the
same as in Fig.~\ref{fig:ratio_flow}).}
\label{fig:ratio_rotation}
\end{figure*}
%-----------------------------------------------------------------------

%

In the present section we estimate this effect in the following simplified way:
\begin{itemize}
\item[-] we assume that starting from the moment of closest approach of
  the two nuclei ($t=0$), the fire streaks rotate for a short time around the geometrical center of the collision, in a way which deflects their trajectory by a given small angle $\alpha$ (Fig.~\ref{drawing-r}). We note that as it will be shown below, the value of the angle $\alpha$ can be constrained by experimental data on directed flow as a function of rapidity.
\item[-] consequently we assume that after the rotation by the angle
  $\alpha$ is complete, the fire streak will follow its modified
  trajectory until the time $t$ in collision c.m.s. reaches the value
  $t=\beta\gamma\tau$, where $\beta$ is the total fire-streak velocity
  and $\gamma=(1-\beta^2)^{-\frac{1}{2}}$. At that moment $t$ pions are
  emitted from the fire streak. As it follows from the above the pion emission point is shifted in the transverse (and also longitudinal) direction with respect to the case with no rotation which we considered in Sec.~\ref{long}. The size of the above shift increases with increasing $\alpha$ and $\tau$.
\item[-] finally, the pion emission occurs following the prescription provided in Sec.~\ref{initial_conditions} but taken along the modified fire-streak trajectory, that is, it is {rotated} by the angle $\alpha$. We assume azimuthal isotropy of pion emission in the rotated frame (which corresponds, for the case with no rotation, to the simulation w/o flow shown in Fig.~\ref{fig:ratio_flow} as blue dashed line). The kinematical variables of the emitted pions in the rotated frame are recalculated back to the original {collision} c.m.s. frame.
\end{itemize}
On the technical level, we simplify the simulation by assuming the
fire-streak trajectory to be directly tilted by the angle $\alpha$ as it is shown in Fig.~\ref{drawing-r} (dashed line). This approximation appears excellent for the values of $\alpha$ considered here: the corresponding shift of the pion emission point is of the order of $\alpha^2$ in the transverse and $\alpha^3$ in the longitudinal direction.

In Fig.~\ref{v1_y_rotation}, we present the result of our simulation for
the initial directed flow of pions (before the action of the EM field),
assuming the rotation of the fire streaks by the angle $\alpha=1.75\deg$, and compared to experimental data on charged pion directed flow from NA49 and WA98 shown before in Fig.~\ref{fig:flow_fit}. As it is evident from the figure, in our model the sidewards deflection of pion emission induced by the rotation of the fire streaks results in rapidity-dependent pion directed flow. This rapidly increases with the angle $\alpha$ resulting from the rotation, thus the experimental data on $v_1(y)$ can be used to constrain the value of the latter angle. We find that rotation by $\alpha=1.75\deg$ gives a good description of the experimental data. We find this fact interesting in itself, as it shows that within our simple approach a connection exists between the measured directed flow and vorticity.

As a consequence the above, in Fig.~\ref{fig:ratio_rotation}
(dash-dotted curve) we present the result of our full simulation made
for the same values of $\tau$ as discussed in Sec.~\ref{long}
(Fig.~\ref{fig:ratio_flow}, blue dashed curve) but including {the}
rotation of fire streaks by a total angle $\alpha=1.75\deg$. Evidently no significant change is visible with respect to the case with no rotation also shown in the figure (dashed curve). This
{is due to}
%we interpret as
%consequence of
the small value of the angle $\alpha$ allowed for our model by the experimental data on $v_1$ (Fig.~\ref{v1_y_rotation}), which also results in
{a small
%very {\it small} values {\it for} the
displacement of the pion emission points.}
% induced by fire-streak rotation.
%Thus,
%we conclude
%From
%our simple study
%the (simplified) estimate
%presented above,
We conclude that for the relatively short pion emission times $\tau$ discussed here, fire-streak vorticity has no much effect on the electromagnetic distortion of $\pi^+/\pi^-$ ratios.
We also note the difference remaining between the result of our
study on fire-streak rotation, and that made in Sec.~\ref{long}
where we applied directed flow from an explicit parametrization of
experimental data as a function of pion rapidity and $p_T$. This we
interpret as originating from the imperfect description of the
$p_T$-dependence of directed flow by our simplified model of
fire-streak rotation, a subject we leave for a future study.

\section{Summary and conclusions}
\label{summary}
%-----------------------

In the present paper we investigated whether the electromagnetic
effects observed in the projectile hemisphere of peripheral $Pb+Pb$
collisions at SPS energies~\cite{NA49_EM} can be described within
the fire-streak model, in its {\em simplified} formulation from
Ref.~\cite{FS_model_1}. This model was shown to describe the
broadening of the pion rapidity distribution as a function of
centrality (or impact parameter). In our opinion the fire-streak
picture provides realistic initial space-time conditions for
quark-gluon plasma creation.

In the fire-streak model the plasma expands in the longitudinal
direction, with its speed depending on the position in the impact
parameter plane. The parts which are close to spectators move with
velocities only slightly smaller than these of the spectators
themselves. Pions are created after some time related to the
hadronization process, which we treated with the help of a free
parameter. After being created charged pions undergo strong
electromagnetic fields generated by fast moving spectators (both
electric and magnetic in the nucleus-nucleus center-of-mass system).

In sequence, we investigated the role of the different contributions
to the observed effect: the transverse positions of the
fire streaks, isospin differences between $\pi^+$ and $\pi^-$
production, the longitudinal evolution of the system and
corresponding pion creation time, and spectator expansion. We have
obtained a satisfactory description of the experimental data for
faster pions ($x_F\geq 0.1$). Rather small pion creation times have
been necessary to describe the data ($0.5<\tau<2$~fm$/c$). These
times are much shorter than claimed from other methods.
We interpret this difference as resulting from the fact that our
study was anchored to high pion rapidities, while the works
summarized in Ref.~\cite{hbt} were dominated by the central rapidity
region of pion production.
A significantly better description of the experimental data was
achieved once spectator expansion was taken into account. The
postulated surface radial velocity of spectator expansion is
$\beta_R=0.2$. The corresponding mean expansion velocity, 0.15$c$ is
reminiscent of results obtained at lower energy~\cite{cugnon81}.

In the present paper we have discussed also the role of the
$v_2$ and $v_1$ flow for the electromagnetically modified
$\pi^+ /\pi^-$ ratio. The $v_2$ and $v_1$ coefficients
have been fitted to the NA49 and WA98
data as a function of the pion rapidity and transverse
momentum. The proposed parametrization has been used then as an
initial condition for the evolution of charged pions in
the electromagnetic field of spectators. While the elliptic flow has
turned out to be unimportant for the $\pi^+ / \pi^-$ ratio,
the directed flow turned out to somewhat reduce the $\pi^+/\pi^-$ ratio
for $x_F >$ 0.25.
%$x_F >$ 0.35.

Finally, we also evaluated the possible influence of transverse
expansion of fire streaks, as well as of their vorticity on our
results for $\pi^+/\pi^-$ ratios. From our simple estimates we concluded
no significant influence of neither effect on the electromagnetic distortion of $\pi^+/\pi^-$
ratios. On the other hand, we have found that rotation of fire streaks
results in the presence of rapidity-dependent directed flow.
Consequently, after the inclusion of the latter rotation we have found that
our model provides a satisfactory description of the experimental data
on pion $v_1$ as a function of rapidity, which can be used to constrain the rotation angle.
A rather small angle, smaller than 2 deg, is needed to describe the
experimental data of NA49 and WA98 collaborations.

To sum up, it was demonstrated in previous works~\cite{FS_model_1,FS_model_2}
that the fire-streak model, in its simplified formulation presented
therein gives a good description of the centrality dependence
of pion rapidity spectra at SPS energies.
At present we conclude that the same model
%provides
%realistic initial conditions for pion production, which
can properly describe the electromagnetic effects on charged pion ratios
at large rapidity.\\

%It was demonstrated in previous works~\cite{FS_model_1,FS_model_2}
%that the fire-streak model, as formulated therein, gives a good
%description of the centrality dependence of pion rapidity spectra at
%SPS energies. At present we conclude that the same model provides
%realistic initial conditions for pion production, which can
%properly describe the electromagnetic effects on charged pion ratios at large rapidity.\\

\section*{Acknowledgments}
%We are indebted to our referee for valuable comments and constructive criticism.

This work was supported by the National Science Centre, Poland under
Grant No. 2014/14/E/ST2/00018.

%------------------------------------------------------------------------------

%------------------------------------------------------------------------------------


\begin{thebibliography}{99}
%
\bibitem{EM_previous_1}
A.~Rybicki and A.~Szczurek, Phys. Rev. C {\bf 75}, 054903 (2007).
%
\bibitem{NA49_EM}
A.~Rybicki, Acta Phys. Polon. B {\bf 42}, 867 (2011), A.~Rybicki
[NA49 Collaboration], PoS EPS{\bf-HEP2009}, 031 (2009).
%
\bibitem{NA49_EM_2}
Similar results have been recently obtained by the NA61/SHINE
Collaboration: K.~Grebieszkow [NA61/SHINE Collaboration], PoS CORFU
{\bf 2018}, 152 (2019), A.~Marcinek [NA61/SHINE Collaboration], Acta
Phys. Polon. B {\bf 50}, no. 6, 1127 (2019). For earlier results,
see also: G.~Ambrosini {\it et al.} [NA52 Collaboration], New Jour.
Phys. {\bf 1}, 23 (1999).
%
\bibitem{hbt}
K. Aamodt {\it et al.} [ALICE Collaboration], Phys. Lett. B {\bf
696}, 328 (2011), see also: C.~Alt {\it et al.} [NA49
Collaboration], Phys. Rev. C {\bf 77}, 064908 (2008).
%
\bibitem{EM_previous_2}
A.~Rybicki and A.~Szczurek, Phys. Rev. C {\bf 87}, 054909 (2013).
%
\bibitem{RS2014}
A.~Rybicki and A.~Szczurek, arXiv:1405.6860 [nucl-th], A.~Rybicki,
A.~Szczurek and M.~Klusek-Gawenda, Acta Phys. Polon. B {\bf 46}, no.
3, 737 (2015).
%
\bibitem{STAR_v1}
L.~Adamczyk {\it et al.} [STAR Collaboration], Phys. Rev. Lett. {\bf
112}, 162301 (2014).
%
\bibitem{FS_model_1}
A.~Szczurek, M.~Kie{\l}bowicz and A.~Rybicki, Phys. Rev. C {\bf 95},
%no.2,
024908 (2017).
%
\bibitem{NA49_rapidity}
T.~Anticic {\it et al.} [NA49 Collaboration], Phys. Rev. C {\bf 86},
054903 (2012).
%%%%%%%%%%%%%%%%
\bibitem{fs1}
R. Hagedorn, Thermodynamics of Strong Interactions, CERN 71-12.
%
\bibitem{fs2}
W.D. Myers, Nucl. Phys. A {\bf 296}, 177 (1978).
%
\bibitem{fs3}
J. Gosset, J.I. Kapusta and G.D. Westfall, Phys. Rev. C {\bf 18},
844 (1978).
%
\bibitem{fs4}
V.K. Magas, L.P. Csernai and D.D. Strottman, Phys. Rev. C {\bf 64},
014901 (2001).
%
\bibitem{fs5}
V.K. Magas, L.P. Csernai and D.D. Strottman, Nucl. Phys. A {\bf
712}, 167 (2002).
%
\bibitem{fs6}
I.N. Mishustin and J.I. Kapusta, Phys. Rev. Lett. 88, 112501 (2002).
%
\bibitem{fs7}
Y.~Xie, D.~Wang and L.~P.~Csernai, Phys. Rev. C {\bf 95}, 031901
(2017).
%
\bibitem{FS_model_2}
A.~Rybicki, A.~Szczurek, M.~Kie{\l}bowicz, A.~Marcinek, V.~Ozvenchuk
and \L{}.~Rozp{\l}ochowski, Phys. Rev. C {\bf 99}, 024908 (2019).
%
\bibitem{UrQMD_1}
S.~Bass {\it et al.}, Prog. Part. Nucl. Phys. {\bf 41}, 255 (1998).
%
\bibitem{UrQMD_2}
M.~Bleicher {\it et al.}, J. Phys. G {\bf 25}, 1859 (1999).
%
\bibitem{fit_1}
R.~Hagedorn, Nuovo Cim. Suppl. {\bf 6}, 311 (1968).
%
\bibitem{fit_2}
W.~Broniowski, W.~Florkowski, and L.~Y.~Glozman, Phys. Rev. D {\bf
70}, 117503 (2004).
%
\bibitem{NA49_flow}
  C.~Alt {\it et al.} [NA49 Collaboration],
  %``Directed and elliptic flow of charged pions and protons in Pb + Pb collisions at 40-A-GeV and 158-A-GeV,''
  Phys.\ Rev.\ C {\bf 68}, 034903 (2003).
%  doi:10.1103/PhysRevC.68.034903
%  [nucl-ex/0303001].
%
\bibitem{WA98_flow}
  H.~Schlagheck [WA98 Collaboration],
  %``Thermalization and flow in 158-GeV/A Pb + Pb collisions,''
  Nucl.\ Phys.\ A {\bf 663}, 725 (2000).
%  doi:10.1016/S0375-9474(99)00703-4
%  [nucl-ex/9909005].
%
\bibitem{isospin}
O.~Chvala {\it et al.}, Eur. Phys. J. C {\bf 33}, S615 (2004).
%
\bibitem{NA49_pp}
C.~Alt {\it et al.} [NA49 Collaboration], Eur. Phys. J. C {\bf 45},
343 (2006).
%
\bibitem{czech}
M.~Czech and A.~Szczurek, J. Phys. G {\bf 32}, 1253 (2006).
%
\bibitem{num49}
NA49 Collaboration, compilation of numerical results,
http://na49info.web.cern.ch/na49info/na49, and references therein.
%
\bibitem{wnm}
A.~Bia\l{}as, M.~Bleszy\'nski and W.~Czy\.z, Nucl. Phys. B~{\bf
111}, 461 (1976).
%
\bibitem{Mazurek}
K.~Mazurek, A.~Szczurek, C.~Schmitt, P.~N.~Nadtochy, Phys. Rev. C
{\bf 97}, no. 2, 024604 (2018).
%
\bibitem{cugnon81}
J. Cugnon and S.~E. Koonin, Nucl. Phys. A {\bf 355}, 477 (1981).
%
\bibitem{na44}
N.~Xu {\it et al.} [NA44 Collaboration], Nucl. Phys. A {\bf 610},
175c (1996).
%
\bibitem{star-nature}
L.~Adamczyk {\it et al.} [STAR Collaboration],
  %``Global $\Lambda$ hyperon polarization in nuclear collisions: evidence for the most vortical fluid,''
  Nature {\bf 548}, 62 (2017).
\end{thebibliography}
\end{document}